\documentclass{ieeeaccess}

\newcommand{\comment}[1]{\ignorespaces}
\usepackage{cite}
\usepackage{color, soul}
\usepackage{graphicx}
\usepackage{hyperref}
\usepackage{changepage}
\usepackage{multirow}
\usepackage{amsmath,amssymb,amsfonts}
\usepackage{mathtools,amssymb}
\usepackage{lipsum,adjustbox}
\usepackage{tikz}
\usepackage{booktabs}
\usepackage{algorithm,algpseudocode}
\usepackage[utf8x]{inputenc}
\usepackage{leftidx}
\usepackage{textcomp}
\usepackage{xcolor}
\usepackage{url}
\usepackage{breqn}
\usepackage{listings}
\usepackage{subcaption}
\usepackage{adjustbox}
\usepackage{textcomp}
\usepackage[english]{babel}
\usepackage{graphicx}
\usepackage{wrapfig}

\usepackage{flushend}

\usepackage[font=scriptsize,labelfont=bf]{caption}
\usepackage{url}
\usepackage{balance}
  \NewSpotColorSpace{PANTONE}
  \AddSpotColor{PANTONE} {PANTONE3015C} {PANTONE\SpotSpace 3015\SpotSpace C} {1 0.3 0 0.2}
  \SetPageColorSpace{PANTONE}%
\usepackage[skins]{tcolorbox}
\newtcolorbox{myframe}[2][]{%
  enhanced,colback=white,colframe=black,coltitle=black,
  sharp corners,boxrule=0.6pt,
  fonttitle=\itshape,
  attach boxed title to top left={yshift=-0.3\baselineskip-0.4pt,xshift=2mm},
  boxed title style={tile,size=minimal,left=0.5mm,right=0.5mm,
    colback=white,before upper=\strut},
  title=#2,#1
}   
\usepackage{booktabs}

\usepackage{pifont}

\newcommand{\xmark}{\text{\ding{55}}}

\usepackage{multirow}
\usepackage{multicol}
\usepackage{breqn}
 \usepackage{pgfplots}
\pgfplotsset{
 compat=1.7,
 every axis/.append style={
 line width=1.25pt,
 tick style={line width=1.25pt, color=black, line cap=round}
 }  
}

\pgfplotsset{width=8cm,compat=1.8}
\usepackage{pgfplotstable}
\usepackage{lineno}

\usepackage{color, colortbl}
\definecolor{YellowOrange}{RGB}{255,127,0}

\definecolor{Gray}{gray}{0.9}
\definecolor{airforceblue}{rgb}{0.36, 0.54, 0.66}
\definecolor{alizarin}{rgb}{0.82, 0.1, 0.26}
\definecolor{amber}{rgb}{1.0, 0.75, 0.0}
\definecolor{amber(sae/ece)}{rgb}{1.0, 0.49, 0.0}
\definecolor{babypink}{rgb}{0.96, 0.76, 0.76}
\definecolor{bronze}{rgb}{0.8, 0.5, 0.2}
\definecolor{battleshipgrey}{rgb}{0.52, 0.52, 0.51}
\definecolor{bole}{rgb}{0.47, 0.27, 0.23}
\definecolor{bulgarianrose}{rgb}{0.28, 0.02, 0.03}
\definecolor{ceil}{rgb}{0.57, 0.63, 0.81}
\definecolor{cerulean}{rgb}{0.0, 0.48, 0.65}
\definecolor{charcoal}{rgb}{0.21, 0.27, 0.31}
\definecolor{coolblack}{rgb}{0.0, 0.18, 0.39}
\definecolor{darkcandyapplered}{rgb}{0.64, 0.0, 0.0}
\definecolor{darkbrown}{rgb}{0.4, 0.26, 0.13}
\definecolor{darkgray}{rgb}{0.66, 0.66, 0.66}
\definecolor{darkjunglegreen}{rgb}{0.1, 0.14, 0.13}
\definecolor{darktaupe}{rgb}{0.28, 0.24, 0.2}
\definecolor{davy\'sgrey}{rgb}{0.33, 0.33, 0.33}
\definecolor{frenchblue}{rgb}{0.0, 0.45, 0.73}
\definecolor{almond}{rgb}{0.94, 0.87, 0.8}
\definecolor{beaublue}{rgb}{0.74, 0.83, 0.9}
\definecolor{beige}{rgb}{0.96, 0.96, 0.86}
\definecolor{bisque}{rgb}{1.0, 0.89, 0.77}
\definecolor{black}{rgb}{0.0, 0.0, 0.0}
\definecolor{fluorescentorange}{rgb}{1.0, 0.75, 0.0}
\definecolor{ghostwhite}{rgb}{0.97, 0.97, 1.0}
\definecolor{antiquewhite}{rgb}{0.98, 0.92, 0.84}
\definecolor{aliceblue}{rgb}{0.94, 0.97, 1.0}

\usepackage{textcomp}
\def\BibTeX{{\rm B\kern-.05em{\sc i\kern-.025em b}\kern-.08em
    T\kern-.1667em\lower.7ex\hbox{E}\kern-.125emX}}
    
\begin{document}
\doi{}


\title{Seeing and Believing: Evaluating the Trustworthiness of Twitter Users}
\vspace{-2mm}
\author{\uppercase{Tanveer Khan}, \uppercase{Antonis Michalas}
	\hfill\href{http://doi.org/10.5281/zenodo.4428240}{\includegraphics[width=2.5\baselineskip,height=3\baselineskip]
		{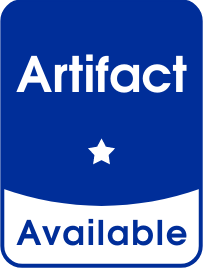}} }
\address{Tampere University, Tampere, Finland \\
(e-mail: \{tanveer.khan, antonios.michalas\}@tuni.fi) }


\vspace{-10mm}
\begin{abstract}
Social networking and micro-blogging services, such as Twitter, play an important role in sharing digital information. Despite the popularity and usefulness of social media, there have been many instances where corrupted users found ways to abuse it, as for instance, through raising or lowering user's credibility. As a result, while social media facilitates an unprecedented ease of access to information, it also introduces a new challenge - that of ascertaining the credibility of shared information. Currently, there is no automated way of determining which news or users are credible and which are not. Hence, establishing a system that can measure the social media user's credibility has become an issue of great importance. Assigning a credibility score to a user has piqued the interest of not only the research community but also most of the big players on both sides - such as Facebook, on the side of industry, and political parties on the societal one. In this work, we created a model which, we hope, will ultimately facilitate and support the increase of trust in the social network communities. Our model collected data and analysed the behaviour of~50,000 politicians on Twitter. Influence score, based on several chosen features, was assigned to each evaluated user. Further, we classified the political Twitter users as either trusted or untrusted using random forest, multilayer perceptron, and support vector machine. An active learning model was used to classify any unlabelled ambiguous records from our dataset. Finally, to measure the performance of the proposed model, we used precision, recall, F1 score, and accuracy as the main evaluation metrics.

\end{abstract}


\titlepgskip=-15pt

\maketitle

\section{\uppercase{Introduction}}
\label{Introduction}
An ever increasing usage and popularity of social media platforms has become the sign of our times -- close to a half of the world's population is connected through social media platforms. The dynamics of communication in all spheres of life has changed. Social media provide a platform through which users can freely share information simultaneously with a significantly larger audience than traditional media. 

As social media became ubiquitous in our daily lives, both its positive and negative impacts have become more pronounced. Successive studies have shown that extensive distribution of misinformation can play a significant role in the success or failure of an important event or a cause~\cite{allcott2017social,metzgar2009social}. Barring the dissemination and circulation of misleading information, social networks also provide the mechanisms for corrupted users to perform an extensive range of illegitimate actions such as spam and political astroturfing~\cite{wang2010don,grier2010spam}. As a result, measuring the credibility of both the user and the text itself has become a major issue. In this work, we assign a credibility score to each Twitter user based on certain extracted features.

Twitter is currently one of the most popular social media platforms with an average of~10,000 tweets per second~\cite{alrubaian2017reputation}. Twitter-enabled analytics do not only constitute a valuable source of information but provide an uncomplicated extraction and dissemination of subject specific information for government agencies, businesses, political parties, financial institutions, fundraisers and many others.

In a recent study~\cite{hindman2018disinformation}, 10~million tweets from~700,000 Twitters accounts were examined. The collected accounts were linked to~600 fakes news and conspiracy sites. Surprisingly, authors found that clusters of Twitter accounts are repeatedly linked back to these sites in a coordinated and automated manner. A similar study~\cite{bovet2019influence} showed that~6.6 million fake news tweets were distributed prior to the 2016 US elections. 


Globally, a number of social and political events in the last three years have been marred by an ever-growing presence of misleading information provoking an increasing concern about their impact on society. This concern translated into an immediate need for the design, implementation, and adoption of new systems and algorithms that will have the ability to \textit{measure} the credibility of a source or a piece of news. Notwithstanding, the seemingly unencumbered growth of social media users is continuing\footnote{In 2020, an estimated~3.23 billion people were using social media worldwide, a number projected to increase to almost~3.64 billion in~2024~\cite{SocialMediaPopulationUpdated}.}. Coupled with the growth in user numbers, the generated content is growing exponentially thus producing a body of information where it is becoming increasingly difficult to identify fabricated stories~\cite{al2015new}. Thereupon, we are facing a situation where a compelling number of unverified pieces of information could be misconstrued and ultimately misused. The research in the field is therefore currently focusing on defining the credibility of the tweets and/or assigning scores to users based on the information they have been sharing~\cite{liu2014tweets,canini2011finding,tinati2012identifying,gupta2012evaluating,moens2014mining,rao2010classifying,al2011experimental,uddin2014understanding}.


\subsection{Our Contribution and Differences with Previous Works}
\label{SS:OC}

We would like to draw your attention to the areas in which this work builds on our previous one~\cite{khan2020trust} and where, we believe, it expounds it and offers new insights. In this work we used \textit{additional ML models}, such as Multi-Layer Perceptron (MLP) and Logistic Regression (LR). Since the MLP model outperformed the LR, we only present the findings for the MLP model. For MLP, we performed the experiments for Tanh, ReLU and Logistics. 
Moreover, unlike~\cite{khan2020trust}, where just one evaluation metric, ``Accuracy'', was used to evaluate the model's performance, in this work, here, we measure the model's performance by using four evaluation metrics -- ``Precision'', ``Recall'', ``F1'' score, and ``Accuracy'' (see table~\ref{tab:ModAccu}). Furthermore, we provide the descriptive statistics of the features (see table~\ref{tab:statistics}) as well as their correlation with the target (see figure~\ref{fig:correlation}) and compare our work with other similar works as SybilTrap~\cite{al2018sybiltrap}  (see table~\ref{tab:Comp table}). Finally, we conduct a comparative review of the user characteristics primarily used in the literature so far, and the ones used in our model and provide supplementary information to help with stratifying trusted and untrusted users (see table~\ref{tab:feateng}). 

Our main contribution can be summarized as follows:


\begin{itemize}
	\item First, we gathered a ~50,000 Twitter users dataset where for each user, we built a unique profile with~19 features (discussed in Section~{\ref{sec:Methodology}}). Our dataset included only users whose tweets are public and have non-zero friends and followers. Furthermore, each Twitter user account was classified as either trusted or untrusted by attaching the trusted and untrusted flag based on different features. These features are discussed in detail in Section~{\ref{sec:ActiveLearning}}.
	
	\item We measured the social reputation score (Section~\ref{Methodology:SR}), a sentiment score (Section~\ref{Methodology:CNS}), an h-index score (Section~\ref{Methodology:RI}), tweets credibility (Section~\ref{Methodology:CTS}) and the influence score~(Section \ref{Methodology:RS}) for each of the analyzed Twitter users.
%
	
	\item To classify a large pool of unlabelled data, we used an active learning model -- technique best suited to the situation where the unlabelled data is abundant but manual labelling is expensive~\cite{tong2001support, settles2009active}. In addition, we evaluated the performance of various ML classifiers.
	
\end{itemize}

\medskip

\subsection{Artifacts}
As a way to support open science and reproducible research and allow other researchers to use, test and hopefully extend/enhance our models we make both our datasets as well as the code for our models available through the Zenodo\footnote{\url{http://doi.org/10.5281/zenodo.4428240}} research artifacts portal. This does not violate \href{https://developer.twitter.com/en.html}{Twitter's developer terms}.

We hope that this work will inspire others to further research this problem and simultaneously kick-start a period of greater trust in social media. 

\subsection{Organization}
\label{SS:PO}
The rest of paper is organized as follows: Related work is discussed in Section~\ref{sec:Related Work}, accompanied by a detailed discussion of our proposed approach in Section~\ref{sec:Methodology}. In Section~\ref{sec:ActiveLearning}, the active learning method and the type of classifier used are discussed. The data collection and experimental results are presented in Section~\ref{sec:Evaluation}. Finally, in Section~\ref{sec:Conclusion}, we conclude the paper.

\section{\uppercase{Related Work}}
\label{sec:Related Work}
Twitter is one of the most popular Online-Social-Networks (OSNs). As data aggregator, it provides data that can be used in research of both historical and current events. Twitter, in relation to other popular OSNs, attracts significant attention in the research community due to its open policy on data sharing and distinctive features~\cite{ratkiewicz2011detecting}. Although openness and vulnerability don't necessarily go hand in hand, on a multiple occasions  malicious users misused Twitter's openness and exploited the service (e.g. political astroturfing, spammers sending unsolicited messages, post malicious links, etc.). 

%
In contrast to mounting evidence towards the negative impact of fake news dissemination, so far, only a few techniques for identifying them in social media have been proposed~\cite{ratkiewicz2011detecting, shu2017fake, gupta2014tweetcred, wang2010don, grier2010spam}. 

Among the most popular and promising ones is evaluating Twitter users and assigning them a reputation score. Authors in~\cite{wang2010don} explored the posting of duplicate tweets and pointed that this behaviour, usually not followed by a legitimate user, affects the reputation score. Posting the same tweet several times has a negative effect on the user's overall reputation score. The authors presented research that supports the above by calculating the edit distance to detect duplications between two tweets posted from the same account. 
 
Furthermore, users have used an immense amount of exchanged messages and information on Twitter to  hijack trending topics~\cite{jain2015hashjacker} and send unsolicited messages to legitimate users. 
Additionally, there are Twitter accounts whose only purpose is to artificially boost the popularity of a specific hashtag thus increasing its popularity and eventually making the underlying topic a trend. The BBC investigated an instance where \pounds150 was paid to Twitter users to increase the popularity of a hashtag and promote it into a trend\footnote{\url{https://www.bbc.com/news/blogs-trending-43218939}}.  

In an attempt to address these problems, researchers have used several ways to detect the trustworthiness of tweets and assign an overall rank to users~\cite{gupta2014tweetcred}. Castillo \textit{et al.,}~\cite{castillo2011information} measured the credibility of tweets based on Twitter features by using an automated classification technique. Alex Hai Wang~\cite{wang2010don} used the followers and friends features to calculate the reputation score. Additionally, Saito and Masuda~\cite{kerres2010managing} considered the same metrics while assigning a rank to Twitter users. 
In~\cite{gupta2012twitter}, authors analysed the tweets relevant to Mumbai attacks\footnote{\url{https://www.theguardian.com/world/blog/2011/jul/13/mumbai-blasts}}. 
Their analysis showed most of the information providers were unknown while the reputation of the remaining ones was very low. 
In another study~\cite{gupta2012credibility} that examined the same event, the information retrieval technique and ML algorithm used found that mere 17\% of the tweets were credibly related to the underlying attacks. 

According to Gilani \textit{et al.,}~\cite{gilani2017bots}, when compared to normal users, bots and fake accounts use a large number of external links in their tweets. Hence, analysing other Twitter features such as URL is crucial for correctly evaluating the overall credibility of a user.
Although, Twitter has included tools to filter out such URLs, several masking techniques can effectively bypass Twitter's existing safeguards. 

In this work, we evaluate the users' trustworthiness and credibility~\cite{Michalas:12:StR, Michalas:14:StRM} by analysing a wide range of features (see Table~\ref{tab:FeaturesNotation}). In comparison to similar works in the field, our model explores a number of factors that could be signs of possible malicious behaviours and makes honest, fair, and precise judgements about the users' credibility.
%
\section{Methodology}
\label{sec:Methodology}
\begin {figure*}
\begin{adjustbox}{width=\textwidth}
	\resizebox{20cm}{0.6cm}{%
		\begin{tikzpicture}[baseline,thick,scale=1,
		level distance=25mm,
		text depth=.8em,
		text height=2em,
		level 1/.style={sibling distance=4.3cm},
		level 2/.style={sibling distance=1.2cm},
		every node/.style = {transform shape, align=center}]]
		\node {Twitter User \\ Influence Score}
		child { node {Tweets \\ Credibility}
			child { node {\\Original \\ Content \\ Ratio} }
			child { node {Liked \\ Ratio} } 
			child { node {URL \\ Ratio} } 
			child { node {Hashtag \\ Ratio} }
			child { node {Retweet \\ Ratio} } } 	
		child {node {h-index \\ Score}
			child { node {Retweet \\ h-index} }
			child { node {Liked \\ h-index} }
		}
		child {node {Sentiment \\ Score}
			child { node {Positive \\ Tweets} }
			child { node {Negative \\ Tweets} } 
			child { node {Neutral \\ Tweets} } 
		}	
		child { node {Social \\ Reputation} 
			child { node {Followers \\ Count} }
			child { node {Friends \\ Count} }
			child { node {Statuses \\ Count} }};
		\end{tikzpicture}
	}
\end{adjustbox}
\caption{Twitter Users Influence Score Calculation}\label{fig:ReputationDiagram}
\end{figure*}
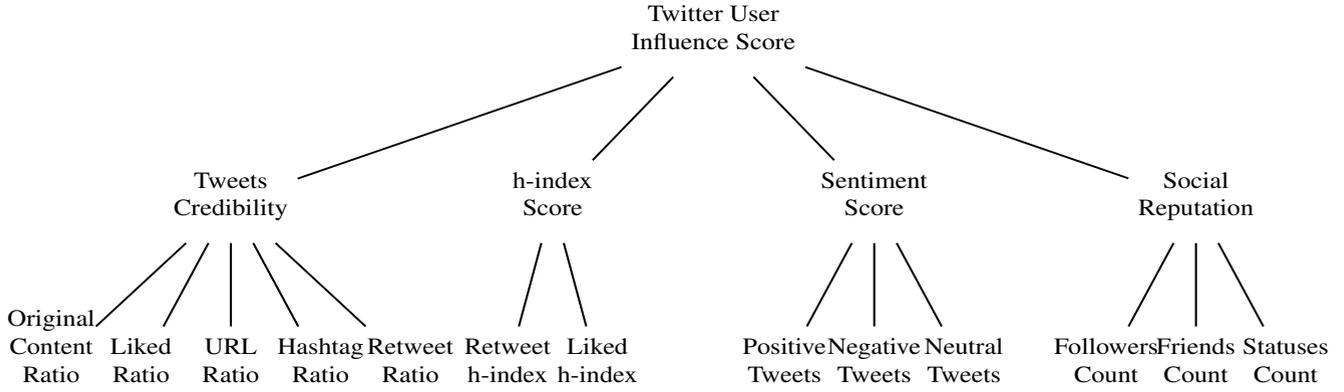

In this section, we discuss the model and main algorithms we used to calculate the user's influence score. Our first goal is to enable the users to identify certain attributes and assess a political Twitter user by considering the influence score that is the outcome of a proper run of our algorithms. Figure~\ref{fig:ReputationDiagram} illustrates the main features we used to calculate users' influence score. We also compare our work with state-of-the-art work in this domain (see Table~\ref{tab:Comp table}). Secondly, the political Twitter users are classified into either trusted or untrusted based on features as social reputation, the credibility of tweets, sentiment score, the h-index score, influential score etc. Accounts containing abusive and/or harassment tweets, low social reputation and h-index score, and low influential score are grouped into untrusted users. The trusted users category envelops more reputable among the users with high h-index score, more credible tweets as well as those having high influential score. We will discuss this in more detail in Section~\ref{sec:ActiveLearning}.

In addition, we also present the approach used to calculate the Twitter users' influence score based on both their context and content features. 
For the user evaluation we took into consideration only the Twitter features that can be extracted through \href{https://pypi.org/project/tweepy/}{Twitter API}. We used the outcome of that evaluation and derived more features to help us provide a better rounded and fair evaluation (Section~\ref{SS:DFTU}). The features, as well as the relevant notation used throughout the paper, are given in Table~\ref{tab:FeaturesNotation}.

\begin{table}[!ht]
	\caption{{Features Considered to Calculate the Influence Score}}
	\centering
	\begin{tabular}{cccc}
		\toprule
		\hline
		\textbf{Notation} &\multicolumn{1}{c} {\textbf{Description}}\\
		\hline
		\midrule
		$N_{fri}(u_i)$: & Number of friends of the user \\
		\midrule
		$N_{fol}(u_i)$: & Number of followers of the user \\
		\midrule
		$N_{ret}$: & Number of retweets for a tweet\\
		\midrule
		$R_{ret}(u_i)$: & Retweet ratio of the user \\
		\midrule
		$N_{lik}$: & Number of likes for a tweet\\
		\midrule		
		$R_{lik}(u_i)$: & Liked ratio of the user\\
		\midrule
		$U_{R}(u_i)$: & Tweet of the user containing URLs\\
		\midrule
		$R_{url}(u_i)$: & URLs ratio of the user\\
		\midrule
		$L(u_i)$: & List count of the user \\
		\midrule
		$N_{T}(u_i)$: & Total number of tweets or Status of the user\\
		\midrule
		$R_{ori}(u_i)$: & Original content ratio of the user\\
		\midrule
		$R_{s}(u_i)$: & Social reputation score of the user \\
		\midrule
		$h_{ind}(u_i)$: & h-index of the user \\
		\midrule
		$R_{hind}(u_i)$: & Retweet h-index of the user \\
		\midrule
		$L_{hind}(u_i)$: & Liked h-index of the user \\
		\midrule
		$Twt_{cr}(u_i)$: & Tweets credibility of the user \\
		\midrule
		$Sen_{s}(u_i)$: & Sentiment score of the user \\
		\midrule
		$N_{neu}(u_i)$: & Neutral tweets\\
		\midrule
		$N_{pos}(u_i)$: & Positive tweets\\
		\midrule
		$N_{neg}(u_i)$: & Negative tweets\\
		\midrule
		$R_{has}(u_i)$: & Hashtag ratio of the user\\
		\midrule
		$Inf(u_i)$: & Influence score\\
		\midrule
		$I_{t}$: & Tweet Index\\
		\hline
		\bottomrule
	\end{tabular}
	\label{tab:FeaturesNotation}
\end{table}

\subsection{Features Selection and Comparison with Previous Models}
\label{SC:PSCM}
The features used for calculating the influence score were based on extensive study of the existing literature. The selected features were used for detection purposes~\cite{fazil2018hybrid, amleshwaram2013cats, yang2013empirical}, assigning a score~\cite{gupta2014tweetcred} or classification purposes~\cite{gilani2017depth}. We used the features given in Table~\ref{tab:FeaturesNotation} to assign an influence score to a $u_i$. Table~\ref{tab:Comp table} provides a comparative overview of existing models based on feature selection.

\begin{table*}[h!]
	\tiny
	\caption{{Models Comparison using Features}}
	\label{tab:Comp table}
	\begin{adjustbox}{max width=\textwidth}
\begin{tabular}{|p{0.45cm}|p{0.8cm}|p{.8cm}|p{.4cm}|p{0.95cm}|p{0.95cm}|p{.79cm}|p{.75cm}|p{.75cm}|p{.8cm}|p{.8cm}|p{.8cm}|p{.65cm}|p{0.9cm}|p{.67cm}|p{.63cm}|p{.4cm}|}
	\hline 
	\rowcolor{Gray}
	Papers & \multicolumn{3}{l|}{$R_{s}(u_i)$}    & \multicolumn{2}{l|}{$h_index$}        & $Sen_{s}(u_i)$ & \multicolumn{6}{l|}{$Twt_{cr}(u_i)$}                                            & \multicolumn{4}{l|}{URLs, List and Mentions}
	\\ \hline 
	
	\rowcolor{Gray}   
	& $N_{fol}(u_i)$ & $N_{fri}(u_i)$ & $N_{T}(u_i)$ & $R_{hind}(u_i)$ & $L_{hind}(u_i)$ &              & $R_{ret}(u_i)$ & $R_{lik}(u_i)$ & $R_{has}(u_i)$ & $R_{url}(u_i)$ & $R_{ori}(u_i)$ & $N_{T}(u_i)$ & $R_{men}(u_i)$ & $N_{M}(u_i)$ & $U_{R}(u_i)$ & $L(u_i)$ \\ \hline
	
	\cite{alrubaian2017reputation}     &   \checkmark  & \checkmark &\xmark&\xmark  &  \xmark &   \checkmark&  \checkmark   & \xmark   &\xmark&\xmark  & \xmark & \checkmark  & \xmark & \checkmark& \checkmark &\xmark\\ \hline
	\cite{al2018sybiltrap}      & \checkmark  &  \checkmark &\xmark          & \xmark                  &        \xmark           &    \xmark&   \checkmark&       \xmark       &  \xmark&  \xmark &\checkmark              &  \checkmark            &  \checkmark  &  \checkmark& \xmark & \xmark       \\ \hline
	\cite{gupta2014tweetcred}  &   \checkmark  & \checkmark &\xmark&\xmark &\xmark&  \checkmark &  \checkmark   &   \checkmark  &\checkmark&  \checkmark&  \checkmark& \xmark &  \checkmark & \checkmark& \checkmark &\xmark\\ \hline
	\cite{fazil2018hybrid}      & \checkmark  &  \checkmark &\xmark          & \xmark                  &        \xmark           &    \checkmark&   \checkmark&       \xmark       &  \checkmark&  \checkmark &\xmark              &  \xmark            &  \checkmark  &  \checkmark& \checkmark & \xmark         \\ \hline
	\cite{amleshwaram2013cats}      &   \checkmark   &\checkmark&    \xmark      &      \xmark             &    \xmark               &  \checkmark &    \xmark          &     \xmark         &     \xmark     &   \xmark       &  \checkmark&   \xmark           & \checkmark & \checkmark &  \checkmark & \xmark         \\ \hline
	\cite{yang2013empirical}     &   \checkmark  & \checkmark & \xmark         &  \xmark                 &   \xmark                &  \checkmark  & \checkmark  & \xmark             &  \checkmark & \checkmark  &    \xmark          &  \checkmark  &       \xmark       &      \xmark    &  \checkmark   &\xmark          \\ \hline
	\cite{gilani2017depth}&   \checkmark  & \checkmark &\xmark &\xmark  &\xmark   &  \checkmark &  \checkmark   &   \checkmark  &\xmark&\xmark  &  \checkmark&  \checkmark & \xmark  & \checkmark& \checkmark &\xmark\\ \hline
	\rowcolor{babypink}
	Proposed     &   \checkmark  & \checkmark &\checkmark & \checkmark &  \checkmark &  \checkmark &  \checkmark   &   \checkmark  &\checkmark&  \checkmark&  \checkmark&  \checkmark &  \checkmark & \checkmark& \checkmark &\checkmark\\ \hline
		\end{tabular}
	\end{adjustbox}
\end{table*}
\subsection{Twitter Features Extraction}
\label{Methodology:TFE}

The pivotal step in the process of assigning a score to a Twitter user is to extract the features linked to their accounts. The features can be either user account specific, such as the number of followers, friends, etc., or user tweet specific, such as the number of likes, retweets, URLs, etc. In our model, we considered both and used them to calculate some additional features. We then combined them all to assign an influence score to a Twitter user. Below we provide more detailed information on features used in our model. 



\paragraph*{Number of Friends}
\label{Methodology:TCFF}
Friend is a user account feature indicating that a Twitter user $(u_i)$ has subscribed to the updates of another $u_i$~\cite{granovetter1977strength}. Following users who are not part of interpersonal ties yields a lot of novel information. One of the important indicators for calculating the $Inf(u_i)$ is the $follower/following$ ratio. The $follower/following$ ratio compares the number of $u_i$'$s$ subscribers to the number of the users, $u_i$ is following. Users are more interested in updates if the $follower/following$ ratio is high~\cite{anger2011measuring}. The ideal $follower/following$ ratio is~1 or close to~1.  In our model, we use the Number of Friends $N_{fri}(u_i)$ as one of the indicators for assigning User's Social Reputation $R_{s}(u_i)$. 
\paragraph*{Number of Followers}
\label{Methodology:TCNF}
$N_{fol}(u_i)$ is another user account feature showing the number of people interested in the specific $u_i$'$s$ tweets. 
As discussed in~\cite{mccoy2017university}, $N_{fol}(u_i)$ is one of the most important parameters for measuring $u_i$'$s$ influence. The more followers a $u_i$ has the more influence he exerts~\cite{leavitt2009influentials}. Preussler \textit{et al.,}~\cite{preussler2010managing} correlates the $N_{fol}(u_i)$ with the reputation of a $u_i$. According to their study, the credibility of a $u_i$ increases as the $N_{fol}(u_i)$ increases. Based on the above we consider the $N_{fol}(u_i)$ an important parameter and use it as input to calculate the $R_{s}(u_i)$.
\paragraph*{Number of Retweets}
\label{Methodology:TCNR}
A tweet is considered important when it receives many positive reactions from other accounts. The reactions may take the form of likes or retweets. Retweets act as a form of endorsement, allowing $u_i$ to forward the content generated by other users, thus raising the content's visibility. It is a way of promoting a topic and is associated with the reputation of the $u_i$~\cite{dutta2018retweet}. Since retweeting is linked to popular topics and directly affects the $u_i$'s reputation, it is a key parameter for identifying possible fake account holders. As described in~\cite{gilani2017bots}, bots or fake accounts depend more on retweets of existing content than posting new ones. In our model, we consider the $N_{ret}$ as one of the main parameters for assigning the $Inf(u_i)$. We calculate the $R_{ret}(u_i)$ (used by \href{https://website.grader.com/}{Twitter grader}) for each tweet by considering $N_{ret}$ divided by $N_{T}(u_i)$, as given in equation~\ref{eq:retweetratio}.

\begin{equation}
R_{ret}(u_i)=\frac{N_{ret}}{N_{T}(u_i)}
\label{eq:retweetratio}
\end{equation}
\paragraph*{Number of Likes}
\label{Methodology:TCL}
The $N_{lik}$ is considered a reasonable proxy for evaluating the quality of a tweet. Authors in~\cite{gilani2017depth} showed that humans receive more likes per tweet when compared to bots. In~\cite{gilani2017classification}, the authors used likes as one of the metrics to classify Twitter accounts as a human user or automated agent. As mentioned in~\cite{alrubaian2017reputation}, if a specific tweet receives a large $N_{lik}$, it can be safely concluded that other $u_i$'$s$  are interested in the tweets of the underlying $u_i$. Based on this observation, we calculate the $R_{lik}(u_i)$ by using the $N_{lik}$ for each tweet and dividing it with $N_{T}(u_i)$ as shown in equation~\ref{eq:likedratio}. 
\begin{equation}
R_{lik}(u_i)=\frac{N_{lik}}{N_{T}(u_i)}
\label{eq:likedratio}
\end{equation}

\paragraph*{URLs}
\label{Methodology:TCU}
URL is a content level feature some $u_i$'$s$ include in their tweets~\cite{hughes2009twitter}. As tweets are limited to a maximum of~280 characters, it is common that $u_i$'$s$ cannot include all relevant information in their tweets. To overcome this issue, $u_i$'$s$ often populate tweets with URLs pointing to a source where more information can be found. In our model, we consider the URL as an independent variable for the engagement measurements~\cite{han2019analysis}. We count the tweets that include a URL and calculate the $R_{url}(u_i)$ by considering the $U_{R}(u_i)$ over the $N_{T}(u_i)$ as given in equation~\ref{eq:urlratio}. 
\begin{equation}
R_{url}(u_i)=\frac{U_{R}(u_i)}{N_{T}(u_i)}
\label{eq:urlratio}
\end{equation}
\paragraph*{Listed Count}
\label{Methodology:LC}
In Twitter, a $u_i$ has the option to form several groups by creating lists of different $u_i$'$s$ (e.g. competitors, followers etc.). Twitter lists are mostly used to keep track of the most influential people\footnote{\url{https://www.postplanner.com/how-to-use-twitter-lists-to-always-be-engaging/}}. The simplest way to measure the $u_i$'s influence is by checking the $L(u_i)$ that the $u_i$ is placed on. Being present in a large number of lists is an indicator that the $u_i$ is considered as important by others. Based on this assumption, we also considered the number of lists that each $u_i$ belongs to.

\paragraph*{Statuses Count}
\label{Methodology:STC}
Compared to the other popular OSNs, Twitter is considered as a service that is \textit{less} social\footnote{\url{https://econsultancy.com/twitter-isn-t-very-social-study/}}. This is mainly due to the large number of inactive $u_i$'$s$ or users who show low motivation in participating in an online discussion. Twitter announced a new feature ``Status availability'', that checks the $N_{T}(u_i)$\footnote{\url{https://www.pocket-lint.com/apps/news/twitter/146714-this-is-what-twitter-s-new-online-indicators-and-status-updates-look-like}}. The status count is an important feature closely related to reporting credibility. If a user is active on Twitter for a longer period, the likelihood of producing more tweets increases, which in turn may affect the author's credibility~\cite{kang2012modeling, ross2016features}. To this end, for the calculation of the $Inf(u_i)$, we also took into account how active users are by measuring how often a $u_i$ performs a new activity\footnote{\url{https://sysomos.com/inside-twitter/most-active-twitter-user-data/}}.
%
\paragraph*{Original Content Ratio}
\label{Methodology:OCR}
It has been observed that instead of posting original content, most $u_i$ retweet posts by others~\cite{anger2011measuring}. As a result, Twitter is changing into a pool of constantly updating information streams. For $u_i$'$s$ with high influence
\ in the network, the best strategy is to use the $30/30/30$ rule: 30\% retweets, 30\% original content, and 30\% engagement~\cite{Soltau2020}. Having this in mind, in our model, we look for $u_i$'s original tweets and add them to their corresponding influence score. We calculate the $R_{ori}(u_i)$ by extracting the retweeted posts by others from the total tweets of $u_i$ as given in equation~\ref{eq:oriconratio}.
\begin{equation}
R_{ori}(u_i)=\frac{N_{T}(u_i)-Retweeted\ other\ tweets}{N_{T}(u_i)}
\label{eq:oriconratio}
\end{equation}
\subsection{Derived Features for Twitter Users}
\label{SS:DFTU}Following the considerations for the selection of the basic features for calculating the $Inf(u_i)$, in this section we elaborate on the extraction of the extra ones. Additionally, we discuss the sentiment analysis technique used to analyse $u_i$'$s$ tweets.

By using the basic features described earlier, we calculated the following features for each $u_i$:

\begin{itemize}
\item Social reputation of a user;
\item Retweet h-index score and liked h-index score;
\item Sentiment score of a user;
\item Credibility of Tweets;
\item Influence score of a user.
\end{itemize}

\paragraph*{User's Social Reputation}
\label{Methodology:SR}
The main factor for calculating the $R_{s}(u_i)$ is the number of users interested in $u_i$'s updates. 
Hence, $R_{s}(u_i)$ is based on the $N_{fol}(u_i)$, $N_{fri}(u_i)$ and $N_{T}(u_i)$~\cite{anger2011measuring, wang2010don}.

\begin{multline}
R_{s}(u_i)=2\log(1+N_{fol}(u_i))+ \\ \log(1+N_{T}(u_i))-\log(1+N_{fri}(u_i))
\label{Eq:SRU}
\end{multline}


In equation~\ref{Eq:SRU} we utilized the log property to make the distribution smoother and minimize the impact of outliers. In addition to that, since $log0$ is undefined, we added~$1$ wherever $log$ appears in equation~\ref{Eq:SRU}. 
In equation~\ref{Eq:SRU}, $R_{s}(u_i)$ is directly proportional to $N_{fol}(u_i)$ and $N_{T}(u_i)$. Based on several studies~\cite{alrubaian2017reputation, anger2011measuring, wang2010don}, $R_{s}(u_i)$ is more dependent on $N_{fol}(u_i)$ hence we give more importance to $N_{fol}(u_i)$ in comparison to $N_{T}(u_i)$ and $N_{fri}(u_i)$. If a $u_i$ has a large $N_{fol}(u_i)$ then the $u_i$ is more reputable. In addition, if a $u_i$ is more active in updating his/her $N_{T}(u_i)$ there are more chances that $u_i$'s tweets receive more likes and get retweeted. While $N_{fol}(u_i)$ and $N_{T}(u_i)$ increase, $R_{s}(u_i)$ also increases and vice versa. Alternatively, if a $u_i$ has less $N_{fol}(u_i)$ in comparison to the $N_{fri}(u_i)$ then, the $R_{s}(u_i)$ is smaller. As can be seen from equation~\ref{Eq:SRU}, there is an inverse relation between $R_{s}(u_i)$ and $N_{fri}(u_i)$. 

\paragraph*{h-Index Score}
\label{Methodology:RI}
The $h_{ind}$ score is most commonly used to measure the productivity and impact of a scholar or scientist in the research community. It is based on the number of publications as well as the number of citations for each publication~\cite{hirsch2005index}. In our work, we use the $h_{ind}$ score for a more accurate calculation of $Inf(u_i)$. 
The $h_{ind}$ of a $u_i$ is calculated considering $N_{lik}$ and $N_{ret}$ for each tweet. To find the $h_{ind}$\footnote{\url{https://gallery.azure.ai/Notebook/Computing-Influence-Score-for-Twitter-Users-1}}, we sort the tweets based on the $N_{lik}$ and $N_{ret}$ (in decreasing order). 

Algorithm~\ref{alg:Rhindex} describes the main steps for calculating the $h_{ind}$ of a $u_i$ based on the \textit{$N_{ret}$}. The same algorithm is used for calculating the $h_{ind}$ of a $u_i$ based on \textit{$N_{lik}$} by replacing $N_{ret}$ with $N_{lik}$.
\begin{algorithm}
\caption{Calculating h-index score based on retweets}\label{alg:Rhindex}
\begin{algorithmic}[1]
	\Procedure{h-Index score}{$h_{ind}$}
	\State Arrange $N_{ret}$ for each tweet of a $u_i$ in decreasing order 
	\For{\texttt{$I_{t}$ in list:}}
	\If{$N_{ret}$ of a tweet $<$ $I_{t}$}
	\State return $I_{t}$
	\EndIf
	\EndFor
	\State return $N_{ret}$
	\EndProcedure
\end{algorithmic}
\end{algorithm}
$R_{hind}(u_i)$ and $L_{hind}(u_i)$ are novel features used for measuring the relative importance of a $u_i$. A tweet that has been retweeted many times and liked by many users is considered as attractive for the readers~\cite{alrubaian2017reputation, riquelme2016measuring}. For this reason, we use $R_{hind}(u_i)$ and $L_{hind}(u_i)$ for measuring the $Inf(u_i)$. The higher the $R_{hind}(u_i)$ and $L_{hind}(u_i)$ score of a $u_i$, the higher will be the $Inf(u_i)$. 

\paragraph*{Twitter User Credibility}
\label{Methodology:credibiltytwitteruser}
The credibility is actually the believability~\cite{castillo2011information} -- that is, providing reasonable grounds for being believed. The credibility of a $u_i$ can be assessed by using the information available on the Twitter platform. In our approach, we use both the $Sen_{s}(u_i)$ and $Twt_{cr}(u_i)$ to find a credible $u_i$. 

\smallskip

\underline{Sentiment Score}:
\label{Methodology:CNS}
It has been observed that OSNs are a breeding ground for the distribution of fake news. In  many cases even a single Twitter post significantly impacted~\cite{wolfsfeld2013social} and affected the outcome of an event. 

Having this in mind, we used \href{https://www.earthdatascience.org/courses/earth-analytics-python/using-apis-natural-language-processing-twitter/analyze-tweet-sentiments-in-python/}{sentiment analysis} and the TextBlob~\cite{loria2014textblob} library, to analyze tweets with the main aim to identify certain patterns that could facilitate identification of credible news.
The sentiment analysis returns a score using polarity values ranging from 1 to -1 and helps in tweet classification. We classified the collected tweets as \textit{(1)} Positive \textit{(2)} Neutral, and \textit{(3)} Negative based on the number of positive, neutral and negative words in a tweet. According to Morozov~\textit{et al.,}~\cite{morozov2014analysing}, the least credible tweets have more negative sentiment words and opinions and are associated with negative social events, while credible tweets, have more positive ones. Hence we classified positive tweets as being the most credible followed by the neutral, and finally the least credible negative tweets. 

Following the tweets classification we assign a $Sen_{s}(u_i)$ to each $u_i$~\cite{alrubaian2017reputation} using the following equation:  

\begin{equation}
Sen_{s}(u_i)=\frac{\sum N_{neu}(u_i)+ \sum N_{pos}(u_i)}{\sum N_{neu}(u_i)+ \sum N_{pos}(u_i) + \sum N_{neg}(u_i)}
\label{eq:NT}
\end{equation}

\smallskip

\underline{Tweets Credibility}:
\label{Methodology:CTS}
Donovan~\cite{odonovan2012credibility} focused on finding the most suitable indicators for credibility. According to their findings, prime indicators for a tweet's credibility are mentions, URLs, tweet length and retweets. Gupta \textit{et al.,}~\cite{gupta2012credibility} ranked tweets based on tweets credibility. The parameters used as an input for the ranking algorithm were: tweets, retweets, total unique users, trending topics, tweets with URLs, start and end date. Based on the existing literature, we compute the $Twt_{cr}(u_i)$ by considering $R_{ret}(u_i)$, $R_{lik}(u_i)$, $R_{has}(u_i)$, $R_{url}(u_i)$ and $R_{ori}(u_i)$ (see equation~\ref{eq:TC}):

\begin{equation} \label{eq:TC}
\resizebox{0.97\hsize}{!}{$%
{Twt_{cr}(u_i) = \left( \frac{R_{ret}(u_i)+R_{lik}(u_i)+R_{has}(u_i)+R_{url}(u_i)}{4} \right) \cdot R_{ori}(u_i)}
$%
}
\end{equation}

To begin, we consider the $R_{ori}(u_i)$ (tweet) by a $u_i$ and for each $R_{ori}(u_i)$ we collect $R_{ret}(u_i)$, $R_{lik}(u_i)$, $R_{has}(u_i)$ and $R_{url}(u_i)$. These four features are linked with the $R_{ori}(u_i)$ such as $R_{ret}(u_i)$ and $R_{lik}(u_i)$ specify the number of times the $R_{ori}(u_i)$ has been retweeted and liked while $R_{has}(u_i)$ and $R_{url}(u_i)$ return only  $R_{ori}(u_i)$ having URLs and hashtags. Hence, to calculate the credibility of tweets, we first calculate the average of these four parameters and then multiply it with $R_{ori}(u_i)$.

\subsection{Influence Score}
\label{Methodology:RS}
The $Inf(u_i)$ is calculated based on the evaluation of \textit{both} content and context features. More precisely, we consider the following features described earlier: 
$R_{s}(u_i)$, $Sen_{s}(u_i)$, $Twt_{cr}(u_i)$ and $h_{ind}(u_i)$. 
After calculating the values of all of these features we use them as input to Algorithm~\ref{alg:reputation}~line~7 which calculates the $Inf(u_i)$. 

\textit{Equation Formulation:}
\label{SSC:EF}
In order to ascertain how influential a $u_i$ is, researchers have taken into consideration one, two or more of the following characteristics:

\begin{itemize}
\item Social reputation~\cite{garcia2017understanding} and weight-age of his tweets~\cite{alrubaian2017reputation};;
\item Tweets credibility~\cite{odonovan2012credibility, alrubaian2017reputation};
\item His ability to formulate new ideas, as well as his active participation in follow-up events and discussions~\cite{chen2011tweet}.
\end{itemize}

An influential $u_i$ must be highly active (have ideas that impact others' behaviours, able to start new discussions etc.,). Additionally, the tweets must be relevant, credible and highly influential (retweeted and liked by a large number of other $u_i$'$s$). If the tweets of highly influential $u_i$'$s$ are credible and the polarity of their tweets' content is positive, they are considered as highly acknowledged and recognized by the community. In short, for a $u_i$ to be considered influential, we combine the efforts of~\cite{alrubaian2017reputation, odonovan2012credibility, garcia2017understanding, chen2011tweet} and calculate the $Inf(u_i)$ using equation~\ref{eq:inf_sco}.
\begin{equation}
\resizebox{0.97\hsize}{!}{$%
	{Inf(u_i) = \frac{Sen_{s}(u_i)+Twt_{cr}(u_i)+R_{s}(u_i)+R_{hind}(u_i)+L_{hind}(u_i)}{5}
	\label{eq:inf_sco}}
$%
}
\end{equation}

\begin{algorithm}[H]
	\caption{Influence score Calculation}\label{alg:reputation}
	\begin{algorithmic}[1]
		
		\Procedure{Influence Score} {$Inf(u_i)$}\medskip
		\State For $i^{th}$ User \medskip
		\State Calculate $R_{hind}(u_i)$ and $L_{hind}(u_i)$, using Algorithm~\ref{alg:Rhindex} \medskip
		\State Calculate $R_{s}(u_i)$ using equation~\ref{Eq:SRU}\smallskip
		\State Calculate $Sen_{s}(u_i)$  using equation~\ref{eq:NT}\smallskip
		\State Calculate $Twt_{cr}(u_i)$ using equation~\ref{eq:TC}\smallskip
		\State Compute $Inf(u_i)$ using equation~\ref{eq:inf_sco} \smallskip
		\EndProcedure
	\end{algorithmic}
\end{algorithm}

\section{Active Learning and ML Models}
\label{sec:ActiveLearning}

\begin{figure}[!ht]
	\centering
	\includegraphics[width=.9\linewidth]{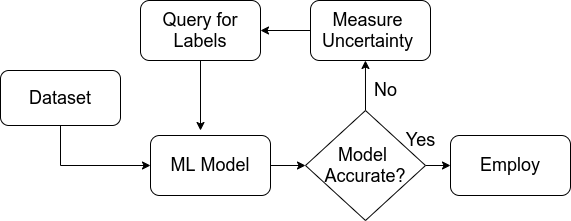}
	\caption{Active Learning Flow }\label{fig:ALF}
\end{figure}

In line with the existing literature, the classification of a $u_i$ is performed on a manually annotated dataset. The manually annotated dataset gives a ground truth, however, manual labelling is an expensive and time-consuming task. In our approach, we used active learning, a semi-supervised ML model that helps in classification when the amount of available labelled data is small. In this model, the classifier is trained using a small amount of training data (labelled instances). Next, the points ambiguous to the classifier in the large pool of unlabelled instances are labelled, and added to the training set~\cite{settles2009active}. This process is repeated until all the ambiguous instances are queried or the model performance does not improve above a certain threshold. The basic flow of active learning approach\footnote{\url{https://github.com/modAL-python/modAL}} is shown in Figure~\ref{fig:ALF}. Based on the proposed model, we first trained our classifier on a small dataset of human-annotated data. Following this step, it then further classified a large pool of unlabelled instances efficiently and accurately.

\begin{figure}[!ht]
	\centering
	\includegraphics[width=1.2\linewidth,scale=1]{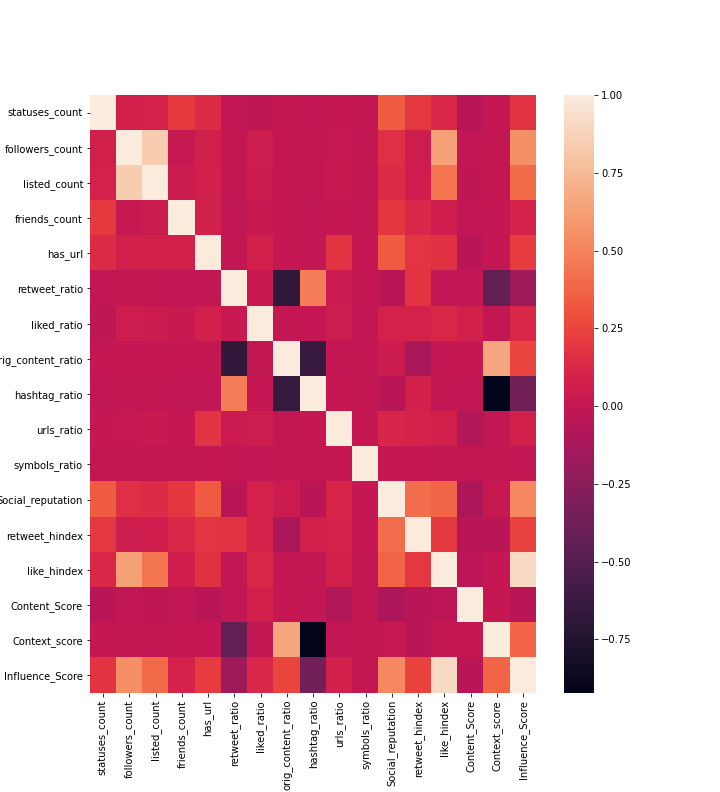}
	\caption{Features Correlation}\label{fig:correlation}
\end{figure}

The steps in our active learning process were as follows:

\begin{itemize}
	
	\item \textbf{Data Gathering:} We gathered unlabelled data for~50,000 $u_i$'$s$. The unlabelled data was then split into a seed -- a small manually labelled dataset consisting of~1000 manually annotated data -- and a large pool of unlabelled data. The seed was then used to train the classifier just like a normal ML model. Using  the seed dataset we classified each political $u_i$ as either trusted or untrusted. 
	
	\item \textbf{Classification of Twitter Users:} Two manual annotators in the field classified~1000 $u_i$'$s$ as trusted or untrusted based on certain features. Out of~1000 $u_i$'$s$,~582 were classified as trusted and the rest~418 as untrusted. For feature selection, we employed the feature engineering technique, and selected the most important features among those presented in Table~\ref{tab:FeaturesNotation}. Based on the existing literature~\cite{daouadi2018organization, pritzkaufinding, ferrara2017disinformation, gurajala2016profile} and correlation among features, certain features were considered the most discriminatory for $u_i$'$s$ classification. We did not include the discriminatory features because they serve as an outlier and are biased. In addition, certain features were distributed almost equally between the trusted and untrusted users, as shown in Table~\ref{tab:feateng}. We discarded both as they do not add any value to classification. However, certain features were good candidates for differentiating trusted and untrusted users such as high $R_{hind}(u_i)$, $L_{hind}(u_i)$,  $Inf(u_i)$, $Sen_{s}(u_i)$, $Twt_{cr}(u_i)$, $R_{s}(u_i)$. In Table~\ref{tab:feateng}, the features marked with $*$ were used for classification in the existing literature~\cite{wang2010don, Myo2018Fake, pritzkaufinding} while the features marked with $\cap$ were based on the correlation among the features. The impact of the individual feature is shown in Figure~\ref{fig:correlation}. The figure indicates that among the features, the $L_{hind}(u_i)$ and $N_{fol}(u_i)$ are very relevant for  assessing $Inf(u_i)$. In addition, all the features except $R_{ret}(u_i)$ and $R_{has}(u_i)$ have a positive impact on the user's $Inf(u_i)$(see Figure~\ref{fig:correlation}).  
	
	\item \textbf{Choosing Unlabelled Instances:} A pool based sampling with a batch size of~100 was used in which~100 ambiguous instances from the unlabelled dataset were labelled and added to a labelled dataset. Different sampling techniques were employed to select the instances from the unlabelled dataset. For the new labelled dataset, the classifier was re-trained and then the next batch of ambiguous unlabelled instances to be labelled was selected. The process was repeated until the model performance did not improve above a certain threshold.
\end{itemize}

\begin{table}[!ht]
	\caption{{Feature Engineering: All values greater than or equal to~0.5 are considered high, whereas those below~0.5 are considered low.}}
	\label{tab:feateng}
	  \begin{adjustbox}{width=.5\textwidth}
	\begin{tabular}{|l|l|l|l|l|}
		\hline
		\rowcolor{Gray}
		\begin{tabular}[c]{@{}l@{}}Discriminatory \\ Features\end{tabular}                           & \begin{tabular}[c]{@{}l@{}}Equally\\ Distributed\end{tabular}              & \begin{tabular}[c]{@{}l@{}}Potential \\ Features\end{tabular} & Trusted & 
		Untrusted \\ \hline
		\multirow{10}{*}{\begin{tabular}[c]{@{}l@{}} \textbullet\ 	$\leftidx{^\cap}{N_{T}(u_i)}$ \end{tabular}} & \multirow{10}{*}{\begin{tabular}[c]{@{}l@{}} \textbullet\ $\leftidx{^*}{R_{url}(u_i)}$ \\  \textbullet\ $\leftidx{^*}{U_{R}(u_i)}$  \end{tabular}} & \textbullet\ $\leftidx{^*}{N_{fol}(u_i)}$                                                        & High              & Low              \\ \cline{3-5} 
		&                                                                                      & \textbullet\ $\leftidx{^\cap}{L(u_i)}$                                                           &      High        &   Low           \\ \cline{3-5} 
		&                                                                                      & \textbullet\ $\leftidx{^*}{N_{fri}(u_i)}$                                                          &     Low          &  High              \\ \cline{3-5} 
		&                                                                                      & \textbullet\ $\leftidx{^\cap}{R_{s}(u_i)}$                                                      &       High        &  Low            \\ \cline{3-5} 
		&                                                                                      & \textbullet\ $\leftidx{^\cap}{Twt_{cr}(u_i)}$                                                          &   High            &   Low           \\ \cline{3-5} 
		&                                                                                      & \textbullet\ $\leftidx{^\cap}{Inf(u_i)}$                                                      &          High     &   Low           \\ \cline{3-5} 
		&                                                                                      &       \textbullet\ $\leftidx{^\cap}{R_{hind}(u_i)}$                                                   &        High      &   Low             \\ \cline{3-5} 
		&                                                                                      & \textbullet\ $\leftidx{^\cap}{R_{ori}(u_i)}$                                                 &          High     &    Low          \\ \cline{3-5} 
		&                                                                                      & \textbullet\ $\leftidx{^*}{R_{has}(u_i)}$                                                          &  Low            &   High         \\ \cline{3-5} 
		&                                                                                      &     \textbullet\ $\leftidx{^\cap}{L_{hind}(u_i)}$                                                    &     High          &  Low            \\ \cline{3-5} 
		&                                                                                      & \textbullet\ $\leftidx{^*}{R_{lik}(u_i)}$                                                          & High              &     Low         \\ \cline{3-5} 
		&                                                                                      &\textbullet\ $\leftidx{^\cap}{Sen_{s}(u_i)}$                                                            &       High        &   Low           \\ \cline{3-5} 
		&                                                                                      & \textbullet\ $\leftidx{^*}{N_{lik}}$                                                          &      High         &     Low         \\ \cline{3-5} 
		&                                                                                      & \textbullet\ $\leftidx{^*}{N_{ret}}$                                                          &      High         &     Low         \\ 
		\cline{3-5} 
		&                                                                                      & \textbullet\ $\leftidx{^*}{R_{ret}(u_i)}$                                                          &      Low         &     High         \\ \hline
	\end{tabular}
  \end{adjustbox}
\end{table}

Among unlabelled instances, active learning finds the most useful ones to be labelled by human annotators. In general, the unlabelled instance which confuses the ML model the most will be the most valuable instance. The following sampling techniques were employed to select instances from the unlabelled dataset\footnote{\url{https://modal-python.readthedocs.io/en/latest/content/query_strategies/uncertainty_sampling.html}}:

\begin{itemize}
	\item \textbf{Uncertainty Sampling:} It is the most common method used to calculate the difference between the most confident prediction and 100\% confidence. 
	
	$U(x)=1-P(\hat{x}|x)$ 
	
	where $\hat{x}$ is the most likely prediction and $x$ is the instance to be predicted. This sampling technique selects the sample with greatest uncertainty.
	
	\item \textbf{Margin Sampling:}  In margin sampling, the probability difference between the first and second most likely prediction is calculated.
	Margin sampling is calculated using equation: 
	
	$M(x)=P(\hat{x_1}|x)-P(\hat{x_2}|x)$, 
	
	where $\hat{x_1}$ and $\hat{x_2}$ are the most likely instances. As the decision is unsure for smaller margins, in this sampling technique, the instance with the smallest margin is selected.

	\item \textbf{Entropy Sampling:} It is the measure of entropy and is defined by the equation:
	
	$H(x)=-\sum_{k}p_k\log(p_k)$
	
	where $p_k$ is the probability of a sample belonging to class $k$. Entropy sampling measures the difference between all the predictions.
\end{itemize} 

\medskip 

Details of the three classifiers we used and their performance characteristics are given below: 
\begin{itemize}
	\item \textbf{Random Forest Classifier (RFC):} An ensemble tree-based learning algorithm~\cite{biau2012analysis} that aggregates the votes from various decision trees to determine the output class of the instance. RFC runs efficiently on large dataset and is capable of handling thousands of input variables. In addition, RFC measures the relative importance of each feature, and produces a highly accurate classifier.
	
	\item \textbf{Support Vector Machine (SVM):} SVM models are commonly used in classification tasks as it  achieves high accuracy with less computation power. The SVM finds a hyperplane in $N$-dimensional space ($N$ represents the number of features) to classify an instance ~\cite{noble2006support}. The goal of SVM is to improve classification accuracy by locating the hyperplane that separates the two classes.
	
	
	\item \textbf{Multilayer Perceptron (MLP):}  A supervised ML algorithm that learns a nonlinear function by training on a dataset. The MLP network is divided into an input layer, hidden layer(s), and output layer~\cite{popescu2009multilayer}. Each layer consist of interconnected neurons transferring information to each other. In our proposed model the MLP consisted of one input and output layer and~50 hidden layers. In addition, the activation functions used in MLP are \textit{Tanh}, \textit{ReLU} and \textit{Logistics}. We do not provide the plots for ReLU activation function as its performance is not as good as Tanh and Logistics (see Table~\ref{tab:ModAccu}).
\end{itemize}

\section{Experimental Results and Model Evaluation}
\label{sec:Evaluation}

\noindent \textbf{Experimental Setup:}
We used Python~3.5 for features extraction and dataset generation. The python script was executed locally on a machine having configuration:
Intel Core i7, 2.80 GHZ,~32GB, Ubuntu~16.04 LTS~64~bit. For training and evaluating the ML models, \href{https://colab.research.google.com/notebooks/welcome.ipynb\#recent=true}{Google Colab} is used. In addition, the modAL framework~\cite{danka2018modal}, an active learning framework for python is used for manually labeling the Twitter users. It is a scikit-learn based platform that is modular, flexible and extensible. We used the pool-based sampling technique for the learner to query the labels of instances, and different sampling techniques for the query strategy. For classification purposes, we used different classifiers, implemented using the scikit-learn library.

\subsection{Dataset and Data Collection}
\label{Eval:DC}  
We used \href{https://pypi.org/project/tweepy/}{tweepy} -- the Twitter's search API for collecting $u_i$'$s$ tweets and features. Tweepy has certain limitations, as it only allows the collection of a certain number of features. Additionally, a data rate cap is in place, which prevents the information collection above a certain threshold.   
Our main concern was to select a sufficient number of users for our dataset. In our dataset, we analysed the Twitter accounts belonging to~50,000 politicians. This dataset was generated in 2020.

The main reason for choosing to evaluate politicians' profiles is their intrinsic potential to influence the public opinion. The content of such tweets originates and exists in the sphere of political life which is, unfortunately, often surrounded by controversial events and outcomes. 
During the selection, we only considered politicians with a \textit{public profile}. Users that seemed to be \textit{inactive} (e.g. limited number of followers and activities) were omitted. In addition, because duplicate data might influence model accuracy, we used the ``max ID'' parameter to exclude them from the data set. Firstly, we requested the most recent tweets from each user (200 tweets at a time) and kept the smallest ID (i.e. the ID of the oldest tweet). Next, we iterate through the tweets and the value of the max ID now will equal the ID of the oldest tweet minus one. This means in the next requests (for tweets collection), we got all the tweets having an ID less than or equal to a specific ID (max ID parameter). For all the subsequent requests, we used the max ID parameter to avoid tweet duplication.

For each $u_i$, we extracted all the features required by our model. Using the extracted features and tweets we calculated $Inf(u_i)$. Furthermore, we collected data that included~19 features including the influence score for~50,000 $u_i$'$s$. Table~\ref{tab:statistics} summarizes the statistics of some of the features examined in the dataset. For features which have no upper bound defined and may have outliers values, such as the number of followers, likes, etc., we used a percentile clip. We then normalized our features using min-max normalization, with~0 being the smallest and~1 being the largest value. 

\begin{table}[]
	\begin{center}
		\caption{Dataset Descriptive Statistics of only Four Features}
		\label{tab:statistics}
	\begin{tabular}{|l|l|l|l|l|}
		\hline
		& \begin{tabular}[c]{@{}l@{}}Status \\ Count\end{tabular} & \begin{tabular}[c]{@{}l@{}}Follower \\ Count\end{tabular} & \begin{tabular}[c]{@{}l@{}}Listed \\ Count\end{tabular} & \begin{tabular}[c]{@{}l@{}}Friends\\ Count\end{tabular} \\ \hline
		Total   & 473152                                                  & 28347960                                                  & 39977                                                   & 451852                                                  \\ \hline
		Mean  & 1112.02                                                 & 5964.66                                                   & 7.14                                                    & 465.04                                                  \\ \hline
		Standard Deviation   & 8174.28                                                 & 199066.10                                                 & 228.97                                                  & 2586.83                                                 \\ \hline

	\end{tabular}
\end{center}
\end{table}

\subsection{Performance measurements of Machine learning and Neural Network Models}
\label{PMML}
We gathered~50,000 unlabelled instances of $u_i$'$s$ and divided our dataset into three subsets: training, testing, and unlabelled data pools. For the training and testing cohorts, we had~1000 manually annotated data instances. The rest of the data was unlabelled (49,000 instances). The model was trained on the labelled training dataset while the performance of the model was measured on the testing dataset.

For the classification, we used different classifiers (all classifiers were trained on the labelled dataset and predictions are reported using~10 fold cross-validation). The \textit{precision}, \textit{recall}, \textit{F1 score} and \textit{accuracy}, were used as the main evaluation metric for the model performance. Precision is the ratio between true positive and all the positives while recall is the ratio of true positive predictions to the total positives examples. F1 score is the weighted average of precision and recall while accuracy measures the percentage of the correctly classified instances. The precision, recall and F1 score are based on true positive, true negative, false positive and false negative. To define these terms, first we considered that the trusted users are positive (labelled as 1), while the untrusted users are negative (labelled as 0). When the model predicts the actual labels, we categorize them as a true positive and true negative, otherwise false positive and false negative. If the model predicts that the user is trusted but the user is not it is false positive, and if the model predicts that the user is untrusted but the user is not then it is a false negative. The performance of the model (precision, recall, and F1 score) was calculated on the testing dataset. To improve the model accuracy, the active learner randomly selected ambiguous data instances from the unlabelled data pool using three different sampling techniques. These ambiguous data instances were then manually labelled by human annotators. The annotated data was added to the labelled dataset. In our model, the human annotators labelled the~100 most ambiguous instances from the unlabelled dataset returned by the active learner. The respective sampling techniques and the accuracy obtained for the top three classifiers (RFC, SVM and MLP) are discussed below.

\begin{figure}[!ht] 
	\captionsetup[subfigure]{font=scriptsize,labelfont=scriptsize}
	\begin{subfigure}[b]{0.45\linewidth}
		\centering
		\includegraphics[width=.9\linewidth]{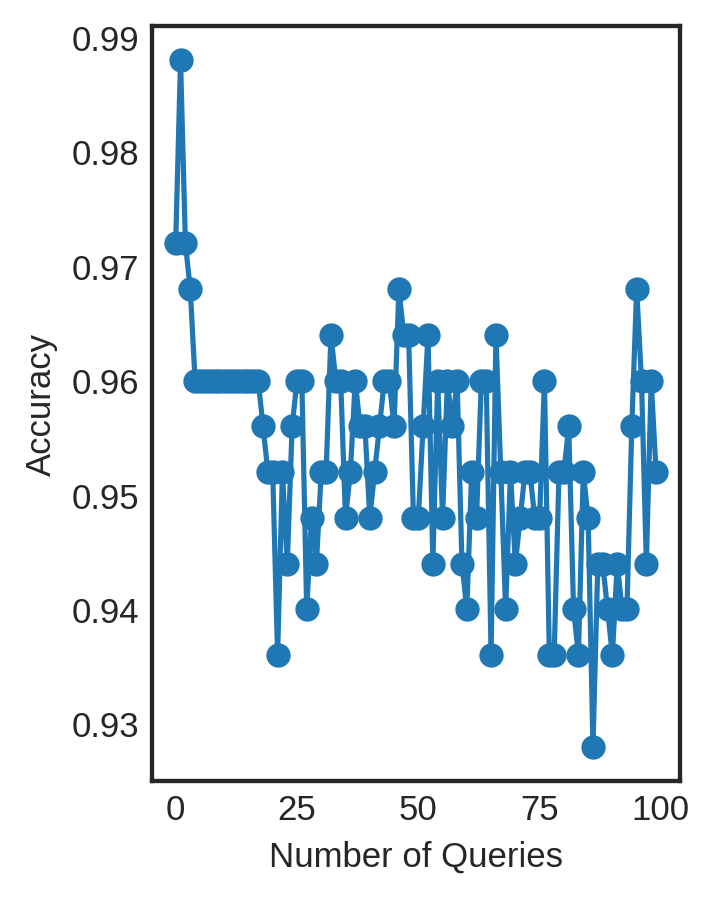} 
		\subcaption{RFC Model}\label{fig:RFCUP}
		\vspace{4ex}
	\end{subfigure}
	\begin{subfigure}[b]{0.45\linewidth}
		\centering
		\includegraphics[width=.9\linewidth]{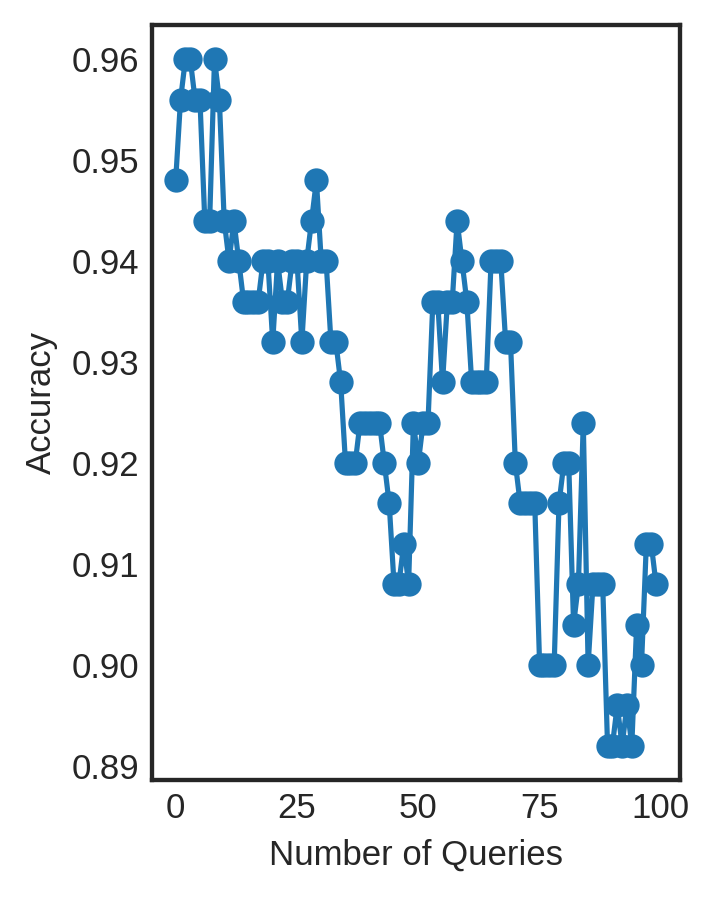}
		\subcaption{SVM Model}\label{fig:SVM}
		\vspace{4ex}
	\end{subfigure} 
	\begin{subfigure}[b]{0.45\linewidth}
		\centering
		\includegraphics[width=.9\linewidth]{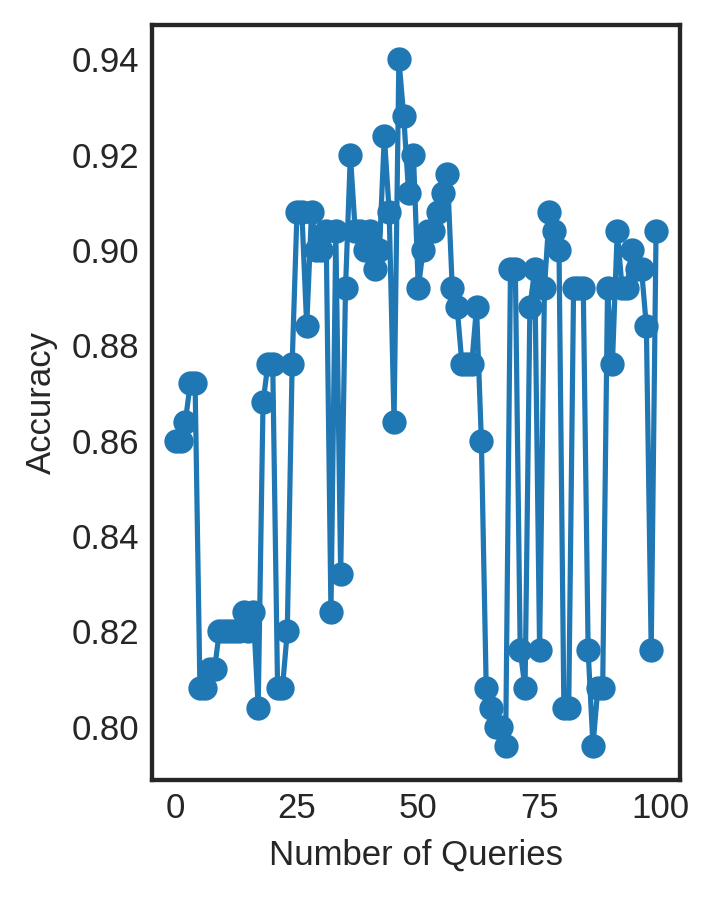}
		\subcaption{MLP Model (Tanh)}\label{fig:LRUS} 
		\vspace{4ex}
	\end{subfigure}
	\begin{subfigure}[b]{0.45\linewidth}
		\centering
		\includegraphics[width=.9\linewidth]{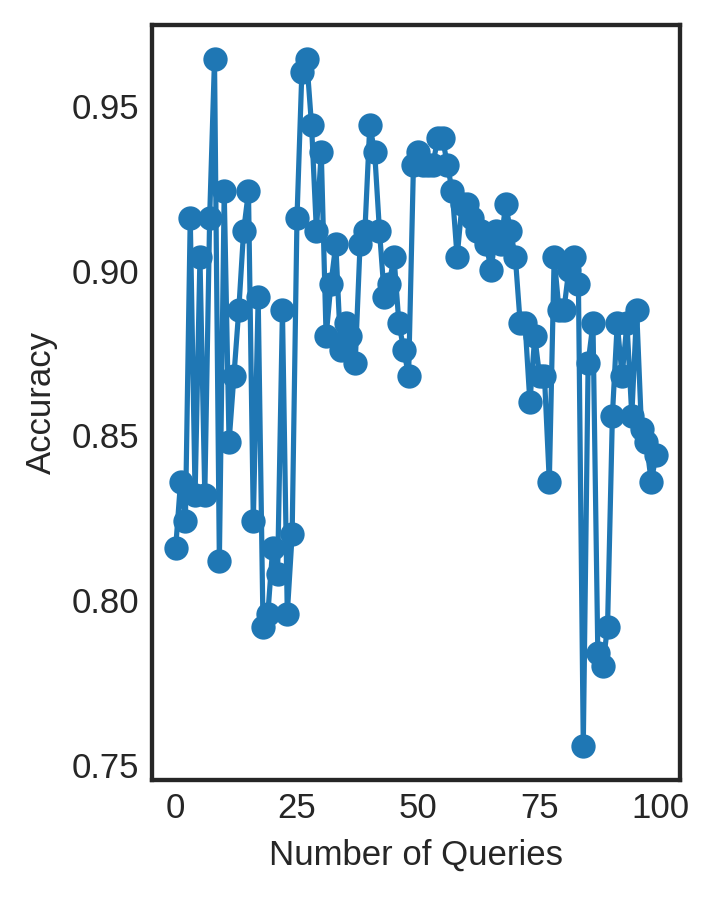}
		\subcaption{MLP Model (Logistics)}\label{fig:MLPLU} 
		\vspace{4ex}
	\end{subfigure} 
	\caption{Accuracy using Uncertainty Sampling}
	\label{fig:accuncertainty}
\end{figure}

\paragraph*{Uncertainty Sampling}
In uncertainty sampling, the least confidence instance is most likely to be considered. In this type of sampling method, the most probable labels are considered and the rest are discarded. The RFC obtained accuracy of~96\%~(Figure \ref{fig:RFCUP}), the SVM obtained an accuracy of~90.8\% (Figure~\ref{fig:SVM}), while the MLP obtained an accuracy of~90\% (Figure~\ref{fig:LRUS}) for Tanh and~84\% for Logistic as given in Figure~\ref{fig:MLPLU}.

%
%

\begin{figure}[!ht] 
		\captionsetup[subfigure]{font=scriptsize,labelfont=scriptsize}
	\begin{subfigure}[b]{0.45\linewidth}
		\centering
		\includegraphics[width=.9\linewidth]{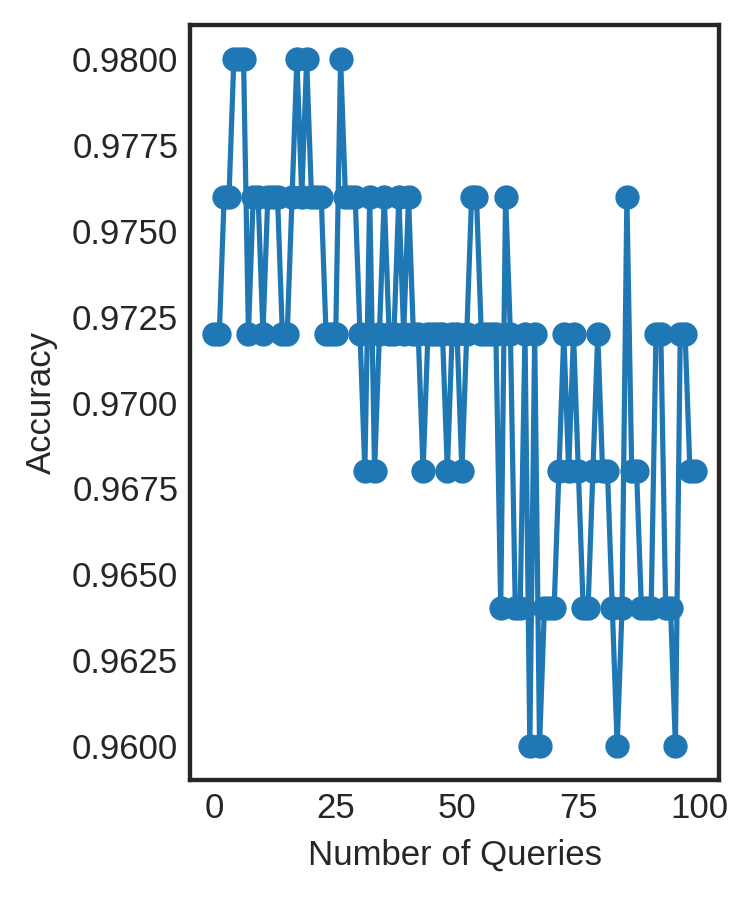} 
		\subcaption{RFC Model}\label{fig:RFCMS}
		\vspace{4ex}
	\end{subfigure}
	\begin{subfigure}[b]{0.45\linewidth}
		\centering
		\includegraphics[width=.9\linewidth]{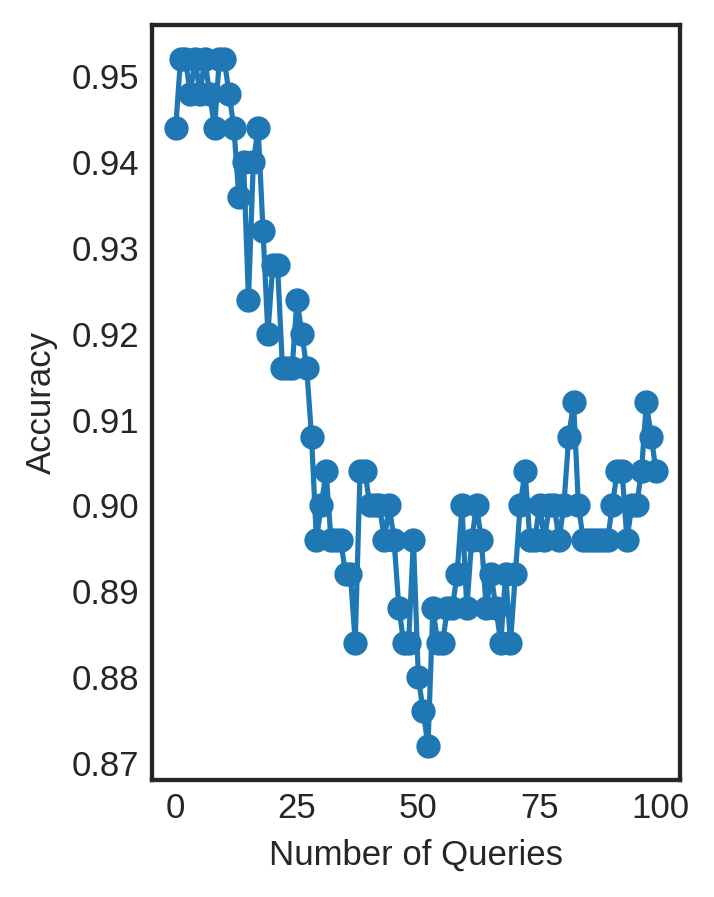} 
		\subcaption{SVM Model}\label{fig:SVCMS}
		\vspace{4ex}
	\end{subfigure} 
	\begin{subfigure}[b]{0.45\linewidth}
		\centering
		\includegraphics[width=.9\linewidth]{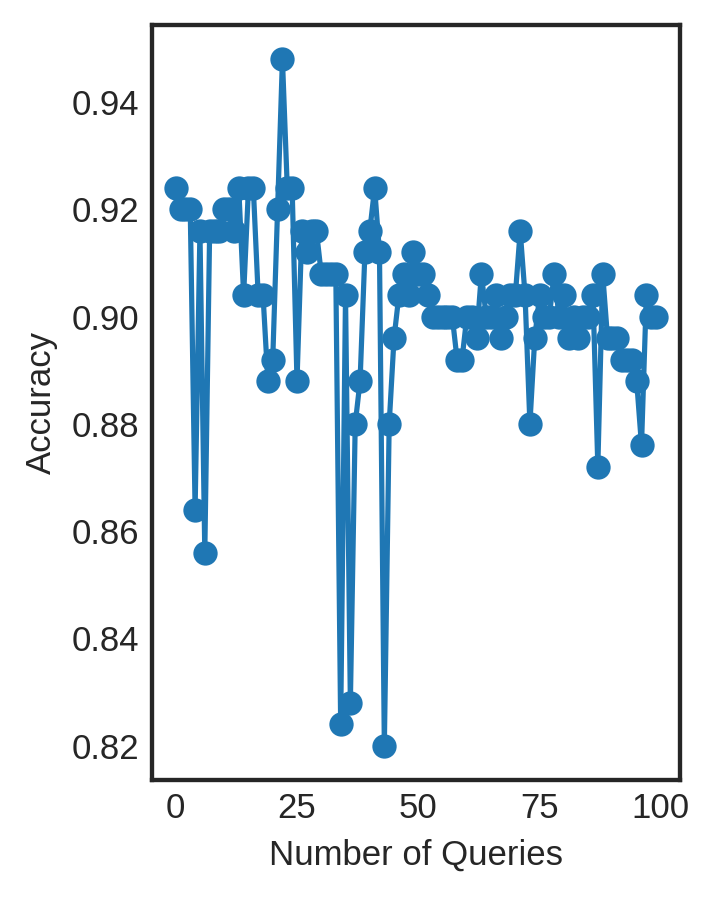} 
		\subcaption{MLP Model (Tanh)}\label{fig:LRMS} 
		\vspace{4ex}
	\end{subfigure}
	\begin{subfigure}[b]{0.45\linewidth}
		\centering
		\includegraphics[width=.9\linewidth]{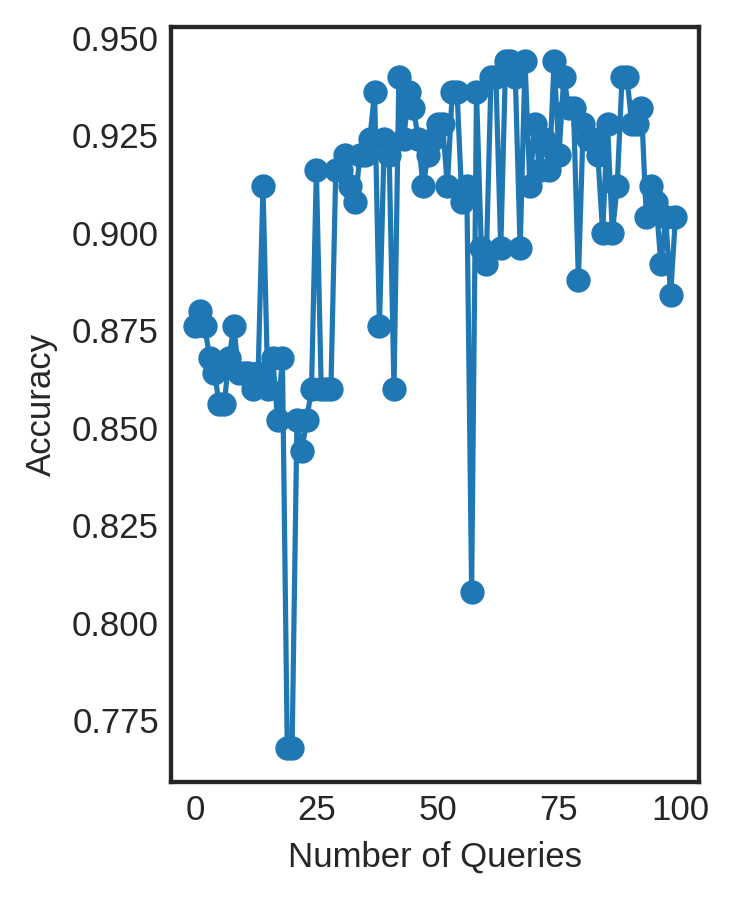} 
		\subcaption{MLP Model (Logistics) }\label{fig:MLPLM} 
		\vspace{4ex}
	\end{subfigure} 
	\caption{Accuracy using Margin Sampling}
	\label{fig:accmargin}
\end{figure}

\paragraph*{Margin Sampling}
In margin sampling, instances with the smallest difference between the first and second most probable labels were considered. The accuracy for RFC, SVM and MLP using margin sampling was~96\%,~91.2\%,~87\% and~88.4\% as shown in Figure~\ref{fig:RFCMS}, Figure~\ref{fig:SVCMS}, Figure~\ref{fig:LRMS} and Figure~\ref{fig:MLPLM} respectively.

\begin{figure}[!ht] 
		\captionsetup[subfigure]{font=scriptsize,labelfont=scriptsize}
	\begin{subfigure}[b]{0.45\linewidth}
		\centering
		\includegraphics[width=.9\linewidth]{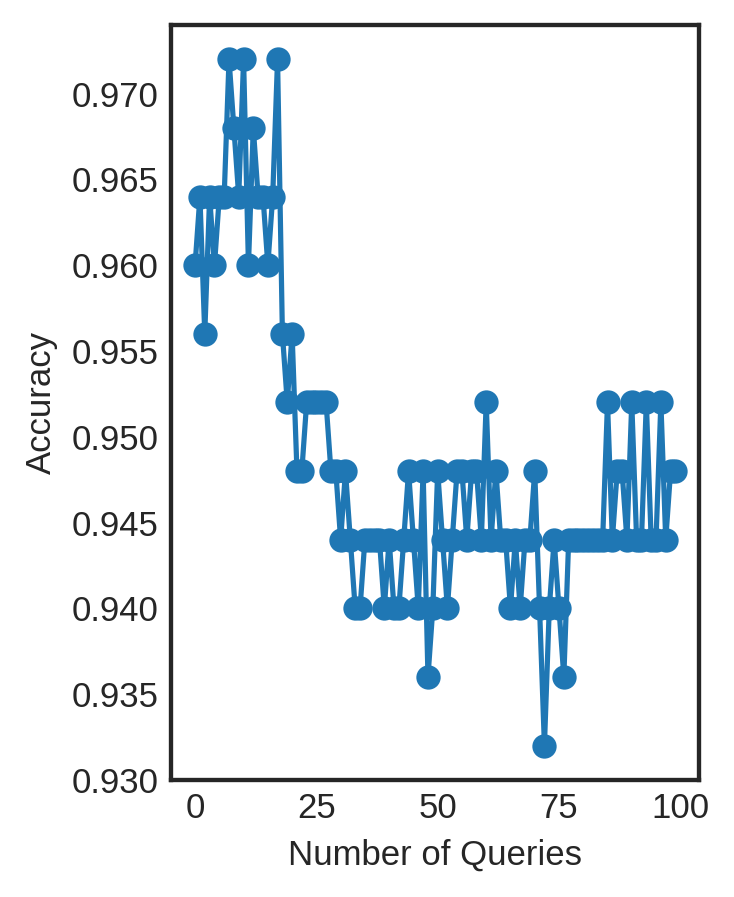}
		\subcaption{RFC Model}\label{fig:RFCES}
		\vspace{4ex}
	\end{subfigure}
	\begin{subfigure}[b]{0.45\linewidth}
		\centering
		\includegraphics[width=.9\linewidth]{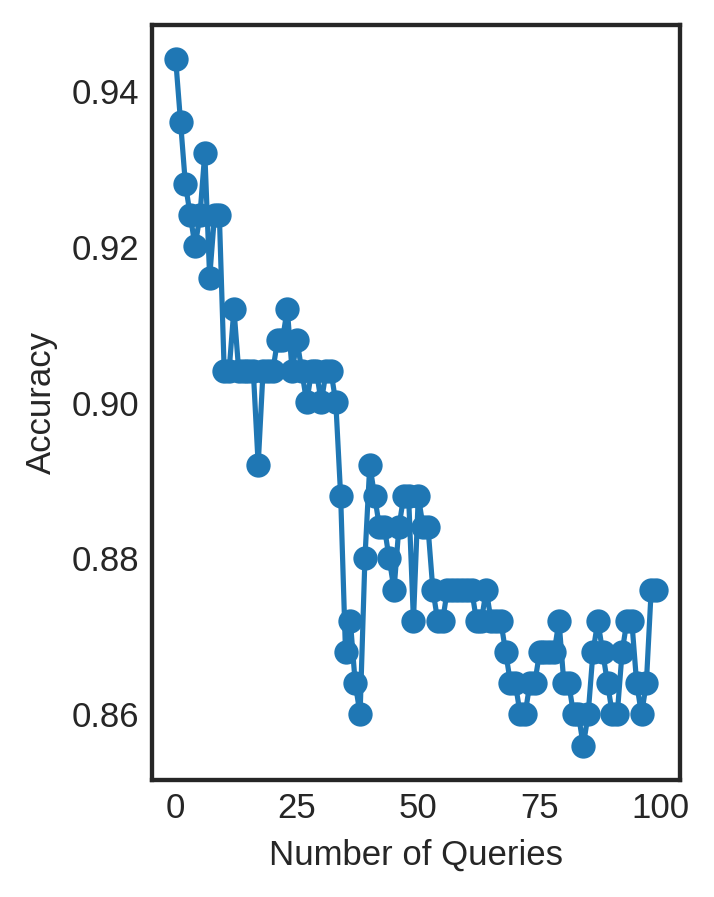}
		\subcaption{SVM Model}\label{fig:SVCES}
		\vspace{4ex}
	\end{subfigure} 
	\begin{subfigure}[b]{0.45\linewidth}
		\centering
		\includegraphics[width=.9\linewidth]{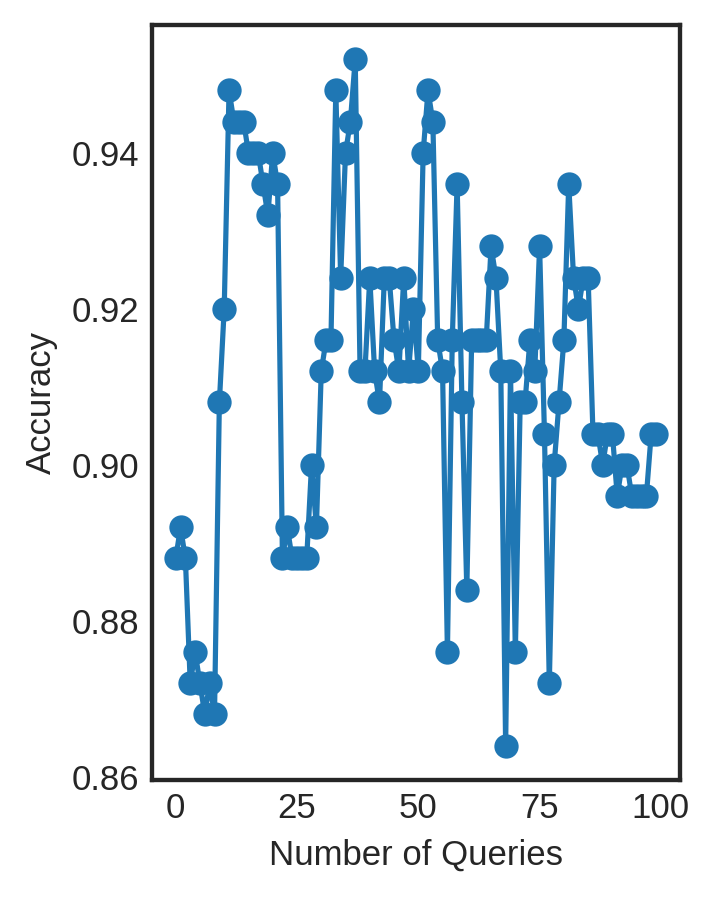}
		\subcaption{MLP Model (Tanh)}\label{fig:LRES} 
		\vspace{4ex}
	\end{subfigure}
	\begin{subfigure}[b]{0.45\linewidth}
		\centering
		\includegraphics[width=.9\linewidth]{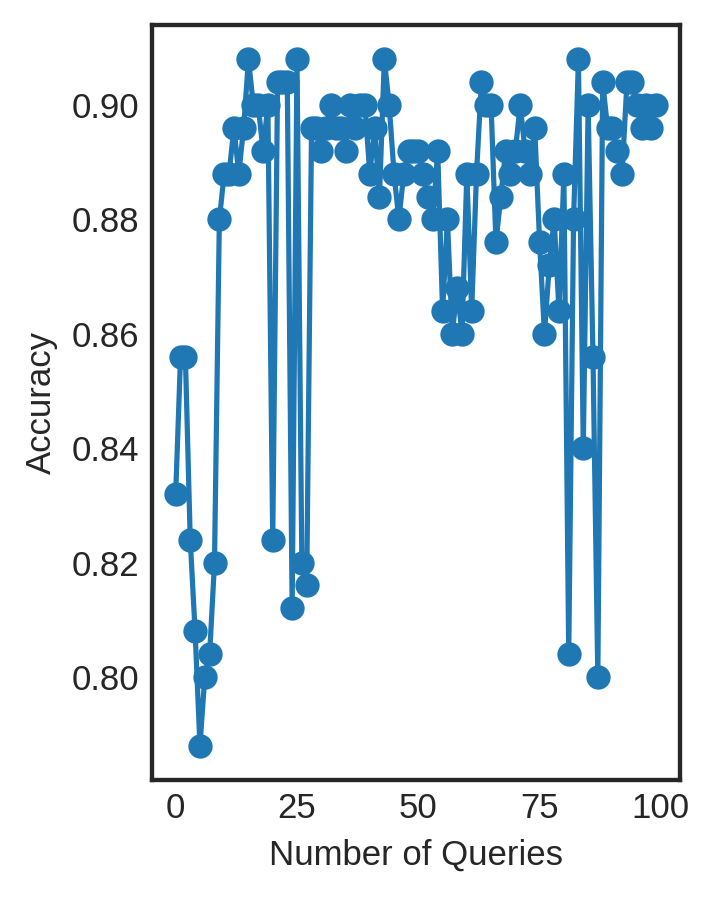}
		\subcaption{MLP Model (Logistics)}\label{fig:MLPLE} 
		\vspace{4ex}
	\end{subfigure} 
		\caption{Accuracy using Entropy Sampling}
		\label{fig:accentropy}
\end{figure}

\paragraph*{Entropy Sampling}
Lastly, the entropy sampling method obtained an accuracy of~95\% for RFC,~88\% for SVM, almost~90\% for MLP (Tanh) and~90\% for MLP (Logistic). Obtained results for the RFC, SVM and MLP, are shown in Figure~\ref{fig:RFCES}, \ref{fig:SVCES},~\ref{fig:LRES} and~\ref{fig:MLPLE}.

Comparison on the performance of our models and different sampling techniques used can be found in Table~\ref{tab:ModAccu}. Precision, recall, F1 score, and accuracy evaluation metrics were used to evaluate the results. Trusted users are represented by~1 while untrusted users are represented by~0 (see Table~\ref{tab:ModAccu}). RFC outperforms the other models in uncertainty sampling, with an F1 score of~96\% for both trusted and untrusted users. Similarly, for margin sampling, RFC received an F1 score of~95\% for untrustworthy users and~97\% for trustworthy users and again outperformed other models. Finally, RFC outperforms in entropy sampling as well, obtaining an F1 score of 95\% for both trusted and untrusted users. Overall, RFC was the best performing algorithm, while MLP (ReLU) had the worst performance. The results obtained by RFC were the best  due to its superior accuracy and better record when it comes to low-dimensional datasets. Similarly, the improved performance, in the case of margin sampling, can be attributed to the fact that it considers the most probable labels probabilities, unlike the other sampling methods. 

\begin{table*}
	\caption{{Comparison of various Models using Different Sampling Techniques}}
	\label{tab:ModAccu}
	\begin{adjustbox}{max width=\textwidth}
\begin{tabular}{|l|l|l|l|l|l|l|l|l|l|l|l|l|l|l|}
	\hline
	\multicolumn{3}{|l|}{\multirow{3}{*}{Models}}                                                                         & \multicolumn{12}{l|}{Sampling Techniques}                                                                                                                                                                                                                                                                  \\ \cline{4-15} 
	\multicolumn{3}{|l|}{}                                                                                                & \multicolumn{4}{l|}{\begin{tabular}[c]{@{}l@{}}Uncertainty\\                Sampling\end{tabular}} & \multicolumn{4}{l|}{\begin{tabular}[c]{@{}l@{}}Margin \\           Sampling\end{tabular}}         & \multicolumn{4}{l|}{\begin{tabular}[c]{@{}l@{}}Entropy \\                  Sampling\end{tabular}} \\ \cline{4-15} 
	\multicolumn{3}{|l|}{}                                                                                                & Precision  & Recall & \begin{tabular}[c]{@{}l@{}}F1 \\ score\end{tabular} & Accuracy               & Precision & Recall & \begin{tabular}[c]{@{}l@{}}F1 \\ score\end{tabular} & Accuracy               & Precision  & Recall & \begin{tabular}[c]{@{}l@{}}F1 \\ score\end{tabular} & Accuracy              \\ \hline
	\multicolumn{2}{|l|}{\multirow{2}{*}{Random Forest}}                                                              & 0 & 0.92       & 0.98   & 0.96                                                & \multirow{2}{*}{0.96}  & 0.94      & 0.97   & 0.95                                                & \multirow{2}{*}{0.96}  & 0.93       & 0.96   & 0.95                                                & \multirow{2}{*}{0.95} \\ \cline{3-6} \cline{8-10} \cline{12-14}
	\multicolumn{2}{|l|}{}                                                                                            & 1 & 0.98       & 0.93   & 0.96                                                &                        & 0.98      & 0.96   & 0.97                                                &                        & 0.96       & 0.94   & 0.95                                                &                       \\ \hline
	\multicolumn{2}{|l|}{\multirow{2}{*}{\begin{tabular}[c]{@{}l@{}}Support Vector \\          Machine\end{tabular}}} & 0 & 0.84       & 0.97   & 0.90                                                & \multirow{2}{*}{0.908} & 0.88      & 0.94   & 0.91                                                & \multirow{2}{*}{0.912} & 0.83       & 0.91   & 0.86                                                & \multirow{2}{*}{0.88} \\ \cline{3-6} \cline{8-10} \cline{12-14}
	\multicolumn{2}{|l|}{}                                                                                            & 1 & 0.98       & 0.86   & 0.92                                                &                        & 0.94      & 0.88   & 0.91                                                &                        & 0.93       & 0.86   & 0.89                                                &                       \\ \hline
	\multirow{6}{*}{\begin{tabular}[c]{@{}l@{}}Multilayer \\  Perceptron\end{tabular}}  & \multirow{2}{*}{Logistics}  & 0 & 0.76       & 0.98   & 0.86                                                & \multirow{2}{*}{0.84}  & 0.83      & 0.89   & 0.86                                                & \multirow{2}{*}{0.884} & 0.95       & 0.83   & 0.89                                                & \multirow{2}{*}{0.90}  \\ \cline{3-6} \cline{8-10} \cline{12-14}
	&                             & 1 & 0.97       & 0.71   & 0.82                                                &                        & 0.92      & 0.88   & 0.90                                                &                        & 0.86       & 0.96   & 0.91                                                &                       \\ \cline{2-15} 
	& \multirow{2}{*}{ReLU}       & 0 & 0.81       & 0.87   & 0.84                                                & \multirow{2}{*}{0.864} & 0.81      & 0.85   & 0.83                                                & \multirow{2}{*}{0.84}  & 0.74       & 0.81   & 0.77                                                & \multirow{2}{*}{0.80} \\ \cline{3-6} \cline{8-10} \cline{12-14}
	&                             & 1 & 0.91       & 0.86   & 0.88                                                &                        & 0.87      & 0.83   & 0.85                                                &                        & 0.85       & 0.80   & 0.83                                                &                       \\ \cline{2-15} 
	& \multirow{2}{*}{Tanh}       & 0 & 0.85       & 0.93   & 0.89                                                & \multirow{2}{*}{0.90}   & 0.89      & 0.81   & 0.84                                                & \multirow{2}{*}{0.87}  & 0.87       & 0.88   & 0.88                                                & \multirow{2}{*}{0.90} \\ \cline{3-6} \cline{8-10} \cline{12-14}
	&                             & 1 & 0.95       & 0.88   & 0.91                                                &                        & 0.86      & 0.92   & 0.89                                                &                        & 0.92       & 0.91   & 0.92                                                &                       \\ \hline
	\end{tabular}
 	\end{adjustbox}
\end{table*}

\section{Conclusion}
\label{sec:Conclusion}
Contemplating the momentous impact unreliable information has on our lives and the intrinsic issue of trust in OSNs, our work focused on finding ways to identify this kind of information and notifying users of the possibility that a specific Twitter user is not credible.


To do so, we designed a model that analyses Twitter users and assigns each a calculated score based on their social profiles, tweets credibility, sentiment score, and h-indexing score. Users with a higher score are not only considered as more influential but also, as having a greater credibility. 
To test our approach, we first generated a dataset of~50,000 Twitter users along with a set of~19 features for each user. Then, we classified the Twitter users into trusted or untrusted using three different classifiers. Further, we employed the active learner approach to label the ambiguous unlabelled instances. During the evaluation of our model, we conducted extensive experiments using three sampling methods. The best results were achieved by using RFC with the margin sampling. We believe this work is an important step towards automating the users' credibility assessment, re-establishing their trust in social networks, and building new bonds of trust between them.

\bibliographystyle{unsrt}

\bibliography{access}

\EOD
\begin{wrapfigure}{l}{2.5cm}
	\includegraphics[width=2.5cm]{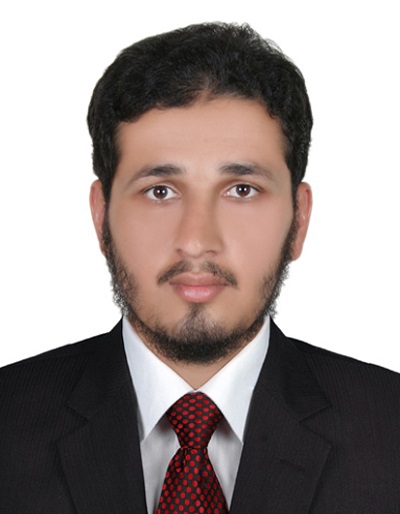}
\end{wrapfigure} 
{Tanveer Khan received the Master Degree in Information Security from COMSATS University 	Islamabad Pakistan. After his Master's, he worked as a Data Analyst on the project CYBER Threat Intelligence Platform at COMSATS University, Islamabad, Pakistan. He also worked as a Junior analyst at Trillium Infosec, Pakistan. Currently, he is working as a Ph.D., Researcher at the Department Computing Sciences, at Tampere University, Finland.  He is also a member of Network and Information Security Group (NISEC) at Tampere University, Finland. His interest is in privacy-preserving machine learning, fake news detection in social networks, cyber security, digital forensics and malware analysis. 
	

	\vspace{0cm}
	
	
	\begin{wrapfigure}{l}{2.5cm}
		\includegraphics[width=2.5cm]{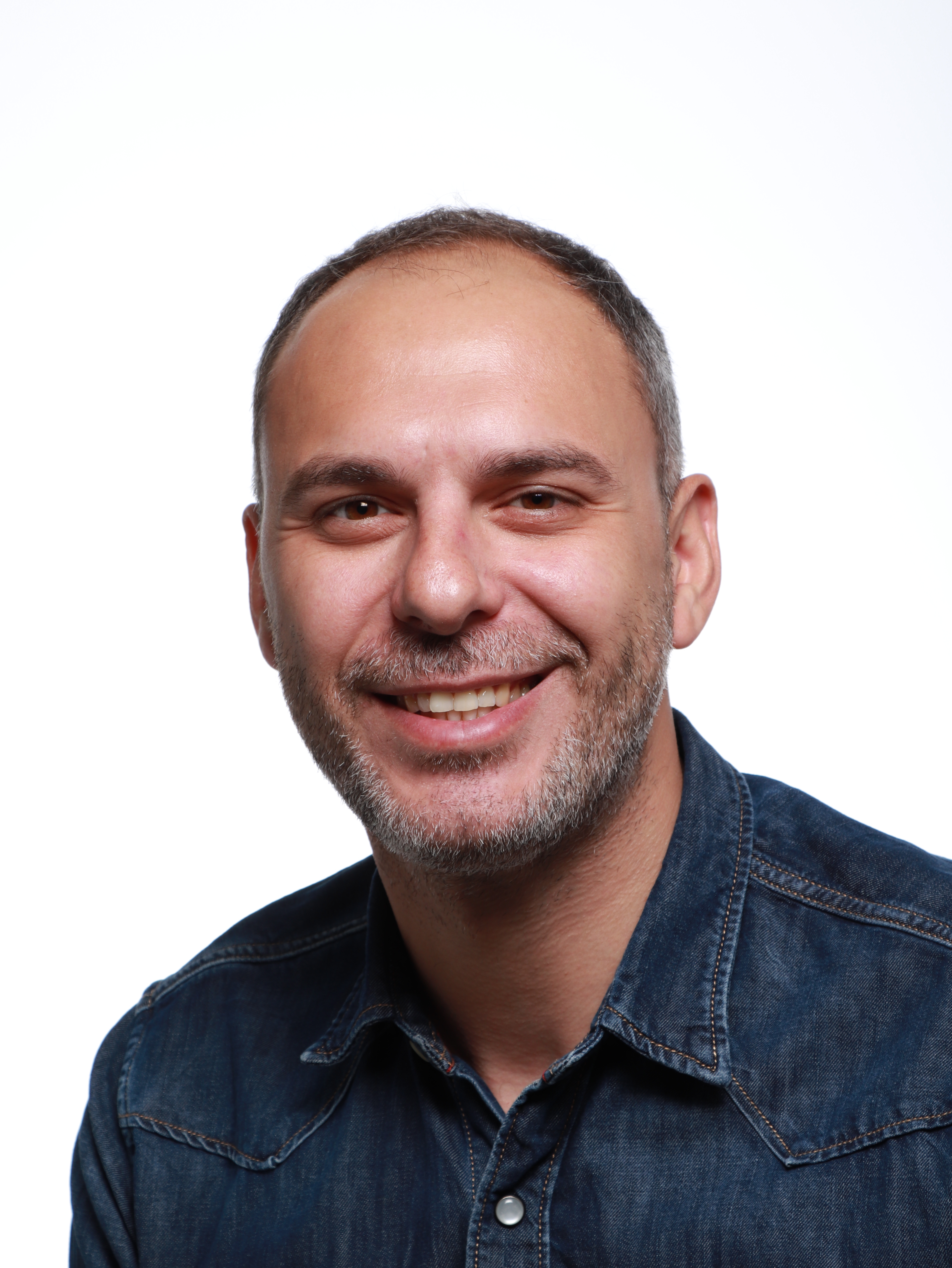}
	\end{wrapfigure} 
	{Prof. Antonis Michalas received his PhD in Network Security from Aalborg University, Denmark and he is currently working as an Assistant Professor at the Department Computing Sciences, at Tampere University, Finland where he also coleads the Network and Information Security Group (NISEC). The group comprises Ph.D., students, professors and researchers. Group members conduct research in areas spanning from the theoretical foundations of cryptography to the design and implementation of leading edge efficient and secure communication protocols. Apart from his research work at NISEC, as an assistant professor he is actively involved in the teaching activities of the University. Finally, his role expands to student supervision and research projects coordination. Furthermore, Antonis has published a significant number of papers in field related journals and conferences and has participated as a speaker in various conferences and workshops. His research interests include private and secure e-voting systems, reputation systems, privacy in decentralized environments, cloud computing, trusted computing and privacy preserving protocols in eHealth and participatory sensing applications.

\end{document}